\documentclass[12pt]{article}
\usepackage{ctex}
\usepackage{amssymb,amsmath,color,amsmath,bm}
\usepackage[colorlinks,linkcolor=blue]{hyperref}
\usepackage{geometry}
\usepackage{tabularx}
\usepackage{bm}
\usepackage{authblk}
\usepackage[square, comma, sort&compress, numbers]{natbib}
\usepackage{multirow}
\usepackage{float}  

\newcommand{\rank}{{\mathrm{rank}}}
\newtheorem{theorem}{Theorem}[section]
\newtheorem{definition}{Definition}[section]
\newtheorem{lemma}{Lemma}[section]
\newtheorem{example}{Example}[section]

\newtheorem{corollary}{Corollary}[section]
\newtheorem{remark}{Remark}[section]

\catcode`@=11 \@addtoreset{equation}{section} \catcode`@=12
\begin{document}
	\title{\bf Four classes of LCD codes from  $(*)$-$(\mathcal{L},\mathcal{P})$-twisted generalized Reed-Solomon codes\thanks{This paper is supported by National Natural Science Foundation of China (Grant No. 12471494) and Natural Science Foundation of Sichuan Province (2024NSFSC2051). The corresponding author is Professor Qunying Liao.}}
	\author{\small Zhonghao Liang}
	\author{\small Qunying Liao
		{\thanks{Z. Liang and Q. Liao are with the College of Mathematical Science, Sichuan Normal University, Chengdu 610066, China (e-mail:liangzhongh0807@163.com;qunyingliao@sicnu.edu.cn)}}
	}
	\affil[] {\small(College of Mathematical Sciences, Sichuan Normal University, Chengdu, 610066, China)}
	\date{}
	\maketitle
	{\bf Abstract.}
	{\small It's well-known that maximum distance separable codes (in short, MDS) and  linear complementary dual (in short, LCD) codes are very important in coding theory and practice. In 2023, Yue et al. \cite{A25} constructed three classes of LCD MDS codes via $(*)$-TGRS codes. Recently, Wu et al. \cite{A27} generalized the results given by Yue et al. and constructed several classes of LCD MDS codes. In this paper, we unify their constructions by defining the  $(*)$-$(\mathcal{L},\mathcal{P})$-
		twisted generalized Reed-Solomon (in short, $(*)$-$(\mathcal{L},\mathcal{P})$-TGRS) code, give the parity-check matrix of $(*)$-$(\mathcal{L},\mathcal{P})$-$\mathrm{TGRS}$ codes, and then construct four classes of LCD codes. Finally, some corresponding examples are given.}\\
	
\section{Introduction}
Let $\mathbb{F}_q$ be the $q$ elements finite field and $\mathbb{F}_{q}^{*}=\mathbb{F}_{q}\backslash\left\{0\right\}$, where $q$ is a prime power. A $k$-dimensional linear subspace of $\mathbb{F}_{q}^{n}$ with minimum (Hamming) distance $d$ is called an $[n, k,d]$ linear code over $\mathbb{F}_{q}$. The dual code of an $[n, k]_{q}$ linear code $\mathcal{C}$ is given by 
\[
\mathcal{C}^{\perp}=\left\{\left(x_{1}, \ldots, x_{n}\right)=\boldsymbol{x}\in\mathbb{F}_q^{n} \mid\langle\boldsymbol{x},\boldsymbol{y}\rangle=\sum\limits_{i=1}^{n} x_{i} y_{i}=0,  \forall \boldsymbol{y}=\left(y_{1}, \ldots, y_{n}\right) \in \mathcal{C}\right\}.
\]

For an $[n, k]_{q}$ linear code $\mathcal{C}$, if $n$ and $k$ are given, then the maximum value of the minimum (Hamming) distance $d$ is determined, i.e, $d\leq n-k+1$. Especially, if $d=n-k+1$, then $\mathcal{C}$ is maximum distance separable (in short, MDS); if $d=n-k$, then $\mathcal{C}$ is almost MDS (in short, AMDS); if $\mathcal{C}^{\perp}$ is also AMDS, then $\mathcal{C}$ is near MDS (in short, NMDS).
 
The generalized Reed-Solomon (in short, GRS) code, as a well-known class of MDS codes, is defined as
$$
\mathrm{GRS}_{k}(\boldsymbol{\alpha})\triangleq\left\{\left(v_{1}f\left(\alpha_{1}\right), \ldots,v_{n}f\left(\alpha_{n}\right)\right) | f(x) \in \mathbb{F}_{q}^{k}[x]\right\},
$$
where $ \boldsymbol{v}=\left(v_{1},\ldots,v_{n}\right)\in\left(\mathbb{F}_{q}^{*}\right)^{n}$, $\boldsymbol{\alpha}=\left(\alpha_{1}, \ldots, \alpha_{n}\right) \in \mathbb{F}_{q}^{n}$ with $\alpha_{i} \neq \alpha_{j}(i \neq j)$, and 
$$\mathbb{F}_{q}^{k}[x]=\left\{f(x)=\sum_{i=0}^{k-1} f_{i} x^{i}\mid f_{i} \in \mathbb{F}_{q}, 0 \leq i \leq k-1\right\}.$$ 
If a linear code $\mathcal{C}$ is not equivalent to any GRS code, then  $\mathcal{C}$ is the non-generalized Reed-
Solomon (in short, non-GRS) type and  $\mathcal{C}$  is called a non-GRS code.

Since MDS codes based on $\mathrm{GRS}$ codes are equivalent to $\mathrm{GRS}$ codes, it's interesting to construct non-GRS MDS codes. Recently, several families of non-GRS MDS codes have been constructed \cite{A1,A2,A3,A4,A5,A6}. In addition, in 2022, Beelen et al.\cite{A8} constructed some non-GRS MDS codes basing on the linear code 
$$
\mathrm{TGRS}_{k}(\boldsymbol{\alpha},\boldsymbol{t},\boldsymbol{h},\boldsymbol{\eta})\triangleq\left\{\left(v_{1}f\left(\alpha_{1}\right), \ldots,v_{n}f\left(\alpha_{n}\right)\right) | f(x) \in \mathcal{P}_{\boldsymbol{t},\boldsymbol{h},\boldsymbol{\eta}}^{n,k}\right\},
$$
where $\boldsymbol{t}=\left(t_{1},\ldots,t_{\ell}\right)\in\left\{1,\ldots,n-k\right\}^{\ell}, \boldsymbol{h}=\left(h_{1},\ldots,h_{\ell}\right)\in\left\{1,\ldots,k-1\right\}^{\ell},\boldsymbol{\eta}=\left(\eta_{1},\ldots,\eta_{\ell}\right)\in\mathbb{F}_{q}^{\ell}, \boldsymbol{v}=\left(v_{1},\ldots,v_{n}\right)\in\left(\mathbb{F}_{q}^{*}\right)^{n}$,   $\boldsymbol{\alpha}=\left(\alpha_{1}, \ldots, \alpha_{n}\right) \in \mathbb{F}_{q}^{n}$ with $\alpha_{i} \neq \alpha_{j}(i \neq j)$, and
\begin{equation}\label{Pnkth}
\mathcal{P}_{\boldsymbol{t},\boldsymbol{h},\boldsymbol{\eta}}^{n,k}=\left\{f(x)=\sum\limits_{i=0}^{k-1} f_{i} x^{i}+\sum\limits_{j=1}^{\ell}\eta_{j} f_{h_{j}} x^{k-1+t_{j}}\mid f_{i} \in \mathbb{F}_{q}\right\}.
\end{equation}
And the $\mathrm{TGRS}_{k}(\boldsymbol{\alpha},\boldsymbol{t},\boldsymbol{h},\boldsymbol{\eta})$ code is called the TGRS code.  Especially, when $\ell=1$ and $\left(\boldsymbol{h},\boldsymbol{t}\right)=(1,0)$, the $\mathrm{TGRS}_{k}(\boldsymbol{\alpha},\boldsymbol{t},\boldsymbol{h},\boldsymbol{\eta})$ code is just the  $(*)$-twisted code given in \cite{A7}. 
In recent years, to construct MDS, self dual or NMDS codes based on $\mathrm{TGRS}_{k}(\boldsymbol{\alpha},\boldsymbol{t},\boldsymbol{h},\boldsymbol{\eta})$ codes have attracted much attention  \cite{A9,A10,A11,A12,A13,A14}. Recently, Zhao et al. \cite{A15} constructed some new MDS TGRS codes by promoting $\mathcal{P}_{\boldsymbol{t},\boldsymbol{h},\boldsymbol{\eta}}^{n,k}$ given in (\ref{Pnkth}). Subsequently, Hu et al.\cite{A16} proposed a
more precise definition for the TGRS code than that given in \cite{A15},  i.e., $$
(\mathcal{L},\mathcal{P})\text{-}\mathrm{TGRS}_{k}(\mathcal{L},\mathcal{P},\boldsymbol{B})\triangleq\left\{\left(v_{1}f\left(\alpha_{1}\right), \ldots,v_{n}f\left(\alpha_{n}\right)\right) | f(x) \in \mathcal{F}_{n,k}(\mathcal{L}, \mathcal{P}, \boldsymbol{B})\right\},
$$
where $\mathcal{L}\in\left\{0,1,\ldots,n-k-1\right\}, \mathcal{P}\in\left\{0,1,\ldots,k-1\right\},\boldsymbol{B}=(b_{i,j})\in\mathbb{F}_{q}^{k\times (n-k)}(0\leq i\leq k-1, 0\leq j\leq n-k-1), \boldsymbol{v}=\left(v_{1},\ldots,v_{n}\right)\in\left(\mathbb{F}_{q}^{*}\right)^{n}$ and 
$$\mathcal{F}_{n,k}(\mathcal{L}, \mathcal{P}, \boldsymbol{B})=\left\{\sum_{i=0}^{k-1} f_{i} x^{i}+\sum_{i \in \mathcal{P}} f_{i} \sum_{j \in \mathcal{L}} b_{i, j} x^{k+j}: f_{i} \in \mathbb{F}_{q}, 0 \leq i \leq k-1\right\}.$$
And the $(\mathcal{L},\mathcal{P})$-$\mathrm{TGRS}_{k}(\mathcal{L},\mathcal{P},\boldsymbol{B})$ code is called the $(\mathcal{L},\mathcal{P})$-TGRS code. At the same time, they gave a necessary and sufficient condition for the $(\mathcal{L},\mathcal{P})$-TGRS code to be MDS, presented a sufficient condition for the $(\mathcal{L},\mathcal{P})$-TGRS code to be NMDS or self-dual, and  discussed the non-GRS property of $(\mathcal{L},\mathcal{P})$-TGRS codes.

For an $[n, k]_{q}$ linear code $\mathcal{C}$, if $\mathcal{C}\bigcap \mathcal{C}^{\perp}=\left\{\boldsymbol{0}\right\}$, then $\mathcal{C}$ is linear complementary dual (in short, LCD). LCD codes have been widely applied in data
storages, communications systems, consumer electronics and
cryptography. In 2024, Carlet et al. \cite{A17} showed that the minimum distance of LCD codes must be as
large as possible. Note that for given $n$ and $k$, if $\mathcal{C}$ is MDS, then the minimum (Hamming) distance of $\mathcal{C}$ is maximum. And so constructing LCD MDS codes is interesting.

Up to now, several families of LCD MDS codes have been constructed \cite{A18,A19,A20,A21,A22,A23,A24,A25,A26,A27} using different methods. Especially, 
in 2023, Yue et al. \cite{A25}  constructed three classes of LCD MDS codes basing on  the following matrix
$$\boldsymbol{G}_{1}=\begin{pmatrix}
	v_{1}\left(1+\eta\alpha_{1}^{k}\right)&\cdots&v_{n}\left(1+\eta\alpha_{n}^{k}\right)\\
	v_{1}\alpha_{1}&\cdots&v_{n}\alpha_{n}\\
	\vdots&\ddots&\vdots\\
	v_{1}\alpha_{1}^{k-1}&\cdots&v_{n}\alpha_{n}^{k-1}\\
\end{pmatrix},
$$ 
where $\eta\in\mathbb{F}_{q}^{*}$. In 2024, Zhu and Liao\cite{A26} also constructed a class of LCD codes basing on the matrix $\boldsymbol{G}_{1}$. Recently, Wu et al.\cite{A27} constructed several classes of LCD MDS codes basing on the following matrix
$$\begin{pmatrix}
	v_{1}\left(1+\eta_{1}\alpha_{1}^{k}+\eta_{2}\alpha_{1}^{k+s}\right)&\cdots&v_{1}\left(1+\eta_{1}\alpha_{n}^{k}+\eta_{2}\alpha_{n}^{k+s}\right)\\
	v_{1}\alpha_{1}&\cdots&v_{n}\alpha_{n}\\
	\vdots&\ddots&\vdots\\
	v_{1}\alpha_{1}^{k-1}&\cdots&v_{n}\alpha_{n}^{k-1}\\
\end{pmatrix}(s=1,2),
$$
respectively. Motivated by the above works, we consider the matrix
$$
	\boldsymbol{G}_{k,*}=\begin{pmatrix}
	v_{1}\left(1+\sum\limits_{t=0}^{\ell}\eta_{t}\alpha_{1}^{k+t}\right)&\cdots&v_{n}\left(1+\sum\limits_{t=0}^{\ell}\eta_{t}\alpha_{n}^{k+t}\right)\\
	v_{1}\alpha_{1}&\cdots&v_{n}\alpha_{n}\\
	\vdots&\ddots&\vdots\\
	v_{1}\alpha_{1}^{k-1}&\cdots&v_{n}\alpha_{n}^{k-1}\\
\end{pmatrix}
$$
and construct four classes of LCD codes.

The paper is organized as follows. In Section 2, we give the definition of the $(*)$-$(\mathcal{L},\mathcal{P})$-$\mathrm{TGRS}$ code and recall some necessary lemmas. In Section 3, we give a parity-check matrix of the $(*)$-$(\mathcal{L},\mathcal{P})$-$\mathrm{TGRS}$ code. In Section 4, we construct four classes of LCD codes basing on $(*)$-$(\mathcal{L},\mathcal{P})$-$\mathrm{TGRS}$ codes. In Section 5, we conclude the whole paper.
\section{Preliminaries}
Throughout this section, for convenience, we fix some notations as the following.
\begin{itemize}
	\item $q$ is an odd prime power.
	\item $\mathbb{F}_q$ is the $q$ elements finite field, and $\mathbb{F}_{q}^{*}=\mathbb{F}_{q}\backslash\left\{0\right\}=\langle\omega\rangle$.
	\item $\mathbb{F}_{q}[x]$ is the polynomial ring over $\mathbb{F}_q$.
	\item $\boldsymbol{\alpha}=\left(\alpha_{1}, \ldots, \alpha_{n}\right) \in \mathbb{F}_{q}^{n}$ with $\alpha_{i} \neq \alpha_{j}(i \neq j)$. 
	\item $ P=\prod\limits_{j=1}^{n} \alpha_{j}=\alpha_{i} P_{i}$.
	\item $u_{i}=\prod\limits_{j=1, j \neq i}^{n}\left(\alpha_{i}-\alpha_{j}\right)^{-1}$ for $1 \leq i \leq n$.
	\item $\boldsymbol{\eta}= \left(\eta_{0}, \ldots, \eta_{\ell}\right) \in\mathbb{F}_{q}^{\ell+1}\backslash\left\{\boldsymbol{0}\right\}$ with $0\leq \ell\leq n-k-1$.
	\item $\boldsymbol{v}=\left(v_{1}, \ldots, v_{n}\right) \in\left(\mathbb{F}_{q}^{*}\right)^{n}$.
\end{itemize}

In this section, basing on Definition 1 in \cite{A16}, we give the definition of $(*)$- $(\mathcal{L},\mathcal{P})$-twisted generalized Reed-Solomon codes  and some necessary lemmas.

\begin{definition}\label{*LPTGRSdefinition} 
Let $n$, $k$ and $\ell$ be integers with $2\leq k\leq n$ and $0\leq \ell\leq n − k-1$. Let  $\boldsymbol{\alpha}=\left(\alpha_{1}, \ldots, \alpha_{n}\right) \in \mathbb{F}_{q}^{n}$ with $\alpha_{i} \neq \alpha_{j}(i \neq j)$, $\boldsymbol{v}=\left(v_{1}, \ldots, v_{n}\right) \in\left(\mathbb{F}_{q}^{*}\right)^{n}$ and $\boldsymbol{\eta}=$ $\left(\eta_{0}, \ldots, \eta_{\ell}\right) \in\mathbb{F}_{q}^{\ell+1}\backslash\left\{\boldsymbol{0}\right\}$. The $(*)$- $(\mathcal{L},\mathcal{P})$-twisted generalized Reed-Solomon (in short, $(*)$- $(\mathcal{L},\mathcal{P})$-TGRS) code is defined as 
$$
(*)\text{-} (\mathcal{L},\mathcal{P})\text{-}\mathrm{TGRS}_{k}(\boldsymbol{\alpha},\boldsymbol{v},\boldsymbol{\eta})\triangleq\left\{\left(v_{1}f\left(\alpha_{1}\right), \ldots,v_{n}f\left(\alpha_{n}\right)\right) | f(x) \in \mathcal{F}_{n,k, \boldsymbol{\eta}}\right\},
$$
where $$\mathcal{F}_{n,k,\boldsymbol{\eta}}=\left\{\sum\limits_{i=0}^{k-1} f_{i} x^{i}+f_{0} \sum\limits_{j=0}^{\ell} \eta_{j} x^{k+j}|f_{i} \in \mathbb{F}_{q}, 0 \leq i \leq k-1\right\}.$$
\end{definition}

\begin{remark}
By Definition \ref{*LPTGRSgeneratormatrix}, it's easy to know that the $(*)$- $(\mathcal{L},\mathcal{P})$-TGRS code has the following generator matrix  
\begin{equation}\label{*LPTGRSgeneratormatrix}
	\boldsymbol{G}_{k,*}=\begin{pmatrix}
		v_{1}\left(1+\sum\limits_{t=0}^{\ell}\eta_{t}\alpha_{1}^{k+t}\right)&\cdots&v_{n}\left(1+\sum\limits_{t=0}^{\ell}\eta_{t}\alpha_{n}^{k+t}\right)\\
		v_{1}\alpha_{1}&\cdots&v_{n}\alpha_{n}\\
		\vdots&\ddots&\vdots\\
		v_{1}\alpha_{1}^{k-1}&\cdots&v_{n}\alpha_{n}^{k-1}\\
	\end{pmatrix}.
\end{equation}
\end{remark}

The following Lemma \ref{LCDequivalent} provides some necessary and sufficient
conditions for a linear code to be LCD.
\begin{lemma}\label{LCDequivalent}
	{\rm(\cite{A27},Proposition 2; \cite{A20}, Lemma 2.7)}	
	Let $\mathcal{C}$ be an $[n, k]$ code over $\mathbb{F}_{q}$ with $k\geq 1$. Suppose that  $\boldsymbol{G}$  and  $\boldsymbol{H}$  are the 
	generator matrix and parity-check matrix for $\mathcal{C}$, respectively. Then, the following statements are equivalent to each other,

$(1)\ \mathcal{C}$ is LCD;

$(2)\ \rank \begin{pmatrix}
		\boldsymbol{G}\\
		\boldsymbol{H}
	\end{pmatrix}=n;$

$(3)$ the $k\times k$ matrix $\boldsymbol{G}\boldsymbol{G}^{T}$ is nonsingular.
\end{lemma}

Next, we recall the definition and the related result of the complete symmetric polynomial. 
\begin{definition}\label{completesymmetricpolynomial} {\rm(\cite{A1}, Lemma 2.6; \cite{A28}, Definition 1.1)}
For any integer $t$, the $t$-th degree complete symmetric polynomial in  $n$-variables is  defined as
$$S_{t}(x_{1},x_{2},\cdots,x_{n})=\begin{cases}
	0,&\text{if}\ t<0;\\
	\sum\limits_{t_{1}+t_{2}+\cdots+t_{n}=t,t_{i}\geq 0}x_{1}^{t_{1}}x_{2}^{t_{2}}\cdots x_{n}^{t_{n}},&\text{if}\ t\geq 0,\\
\end{cases}$$
and denote $S_{t}(x_{1},x_{2},\cdots,x_{n})$ by $S_{t}$.
\end{definition}

\begin{lemma}\label{usapowersum} {\rm(\cite{A1}, Lemma 2.6)}
Let $u_{i}=\prod\limits_{j=1, j \neq i}^{n}\left(\alpha_{i}-\alpha_{j}\right)^{-1}$ for $1 \leq i \leq n$.  Then for any subset $\left\{\alpha_{1}, \ldots, \alpha_{n}\right\}\subseteq\mathbb{F}_{q}$ with $n\geq 3$, we have
$$\sum\limits_{i=1}^{n}u_{i}\alpha_{i}^{h}=\begin{cases}
	0,&\text{if}\ 0\leq h\leq n-2;\\
	S_{h-n+1}(\alpha_{1},\cdots,\alpha_{n}),&\text{if}\ h\geq n-1.\\
\end{cases}$$ 

\end{lemma} 

By the proof of Lemma 2.8 in \cite{A27}, we have the following
\begin{lemma}\label{systemequation}
{\rm(\cite{A27})}Let $\boldsymbol{\alpha}=\left(\alpha_{1}, \ldots, \alpha_{n}\right) \in \mathbb{F}_{q}^{n}$ with $\alpha_{i} \neq \alpha_{j}(i \neq j)$, $P_{i}=\prod\limits_{j \neq i, j=1}^{n} \alpha_{j}$ and  $u_{i}=\prod\limits_{j=1, j \neq i}^{n}\left(\alpha_{i}-\alpha_{j}\right)^{-1}$ for $1 \leq i \leq n$. Then the system of equations over $\mathbb{F}_{q}$
$$\begin{pmatrix}
1&\cdots&1\\
\alpha_{1}&\cdots&\alpha_{n}\\
\vdots&\ddots&\vdots\\
\alpha_{1}^{n-1}&\cdots&\alpha_{n}^{n-1}
\end{pmatrix}\begin{pmatrix}
x_{1}\\
x_{2}\\
\vdots\\
x_{n}
\end{pmatrix}=\begin{pmatrix}
1\\
0\\
\vdots\\
0
\end{pmatrix}$$
has a unique nonzero solution $(w_{1},\cdots,w_{n})^{T},$ where $w_{i}=(-1)^{n-1} u_{i} P_{i}$ $(i=1,2,\ldots,n)$.
\end{lemma}

The following Lemma \ref{mxroot} is crucial for proving our main results.

\begin{lemma}\label{mxroot}
Let $n\mid (q-1)$, $\lambda\in\mathbb{F}_{q}^{*}$ with $\mathrm{ord}(\lambda)\mid \frac{q-1}{n}$, and $\beta_{1},\ldots,\beta_{n}$ be all roots of $m(x)=x^n-\lambda\in\mathbb{F}_{q}[x]$ in $\mathbb{F}_{q^s}$,  where $s$ is an integer with $s\geq 1$. Then $\beta_{i}\in\mathbb{F}_{q}^{*}(1\leq i\leq n)$ and $\beta_{i}\neq \beta_{j}(1\leq i\neq j\leq n)$.
\end{lemma}
{\bf Proof}. Note that $n\mid (q-1)$ and  $\lambda\in\mathbb{F}_{q}^{*}$ with $\mathrm{ord}(\lambda)\mid \frac{q-1}{n}$, thus for $1\leq i\leq n$, we have 
$$\beta_{i}^{q-1}=\left(\beta_{i}^{n}\right)^{\frac{q-1}{n}}=\lambda^{\frac{q-1}{n}}=1,$$ 
it means $\beta_{i}\in \mathbb{F}_{q}^{*}$ for $1\leq i\leq n$. In addition, by $m^{\prime}(x)=nx^{n-1}$, we can get $$(m(x),m^{\prime}(x))=(x^{n}-\lambda,nx^{n-1})=1,$$
it means that $\beta_{i}\neq \beta_{j}(1\leq i\neq j\leq n)$.

From the above, we complete the proof of Lemma $\ref{mxroot}$. 
$\hfill\Box$
\section{The parity-check matrix of the $(*)$-$(\mathcal{L},\mathcal{P})$-$\mathrm{TGRS}$ code}
In this section, for the $(*)$-$(\mathcal{L},\mathcal{P})$-twisted generalized Reed-Solomon code, we give a parity-check matrix as the following  
\begin{theorem}\label{*LPTGRSparitymatrix}
Let  $\boldsymbol{\alpha}=\left(\alpha_{1}, \ldots, \alpha_{n}\right) \in \mathbb{F}_{q}^{n}$ with $\alpha_{i} \neq \alpha_{j}(i \neq j)$, $\boldsymbol{v}=\left(v_{1}, \ldots, v_{n}\right) \in\left(\mathbb{F}_{q}^{*}\right)^{n}$ and $\boldsymbol{\eta}=$ $\left(\eta_{0}, \ldots, \eta_{\ell}\right) \in\mathbb{F}_{q}^{\ell+1}\backslash\left\{\boldsymbol{0}\right\}$. Let   $P=\prod\limits_{j=1}^{n} \alpha_{j}=\alpha_{i}P_{i}$, $u_{i}=\prod\limits_{j=1, j \neq i}^{n}\left(\alpha_{i}-\alpha_{j}\right)^{-1}$ for $1 \leq i \leq n$. Let  $2\leq k\leq n-(\ell +1),0\leq \ell\leq n-k-1$. Then the  $(*)$-$(\mathcal{L},\mathcal{P})$-$\mathrm{TGRS}$ code has the parity-check matrix 
\begin{equation}\label{*LPTGRSparity-checkmatrix}
\boldsymbol{H}_{n-k,*}=\begin{pmatrix}
	\cdots&\frac{u_{j}}{v_{j}}&\cdots\\
	\cdots&\frac{u_{j}}{v_{j}}\alpha_{j}&\cdots\\
	\vdots&\vdots&\vdots\\
	\cdots&\frac{u_{j}}{v_{j}}\alpha_{j}^{n-k-(\ell+2)}&\cdots\\
	\cdots&\frac{u_{j}}{v_{j}}\left(\alpha_{j}^{n-k-(\ell+1)}-(-1)^{n-1}P_{j}\varOmega_{\ell+1}\right)&\cdots\\ 
\vdots&\vdots&\vdots\\
	\cdots&\frac{u_{j}}{v_{j}}\left(\alpha_{j}^{n-k-1}-(-1)^{n-1}P_{j}\varOmega_{1}\right)&\cdots
\end{pmatrix}_{(n-k)\times n},
\end{equation} 
where
$$
\varOmega_{r}=\sum\limits_{t=0}^{\ell}\eta_{t}S_{t+1-r}\left(\alpha_{1},\ldots,\alpha_{n}\right),1\leq r\leq \ell+1.
$$
\end{theorem}
{\bf Proof}. Note that $\boldsymbol{G}_{k,*}$ is the generator matrix of the  $(*)$-$(\mathcal{L},\mathcal{P})$-$\mathrm{TGRS}$ code. To prove that $\boldsymbol{H}_{n-k,*}$ is the parity-check matrix of the  $(*)$-$(\mathcal{L},\mathcal{P})$-$\mathrm{TGRS}$ code, we only need to check that $\mathrm{rank}(\boldsymbol{H}_{n-k,*})=n-k$ and $\boldsymbol{G}_{k,*}\boldsymbol{H}_{n-k,*}^{T}=\boldsymbol{0}$.

Firstly, we prove $\mathrm{rank}(\boldsymbol{H}_{n-k,*})=n-k$. In fact, since the vectors $$(\frac{u_{1}}{v_{1}} ,\cdots,\frac{u_{n}}{v_{n}} ),(\frac{u_{1}}{v_{1}}\alpha_{1},\cdots,\frac{u_{n}}{v_{n}}\alpha_{n}),\cdots,(\frac{u_{1}}{v_{1}}\alpha_{1}^{n-1},\cdots,\frac{u_{n}}{v_{n}}\alpha_{n}^{n-1})$$
are $\mathbb{F}_{q}$-linearly independent, we know that the row vectors of the matrix
$$\boldsymbol{H}=\boldsymbol{H}_{n-k,*}\begin{pmatrix}
\alpha_{1}&&&&\\
&\alpha_{2}&&&\\
&&\ddots&&\\
&&&&\alpha_{n}\\
\end{pmatrix}=\begin{pmatrix}
	\cdots&\frac{u_{j}}{v_{j}}\alpha_{j}&\cdots\\
	\cdots&\frac{u_{j}}{v_{j}}\alpha_{j}^{2}&\cdots\\
	\vdots&\vdots&\vdots\\
	\cdots&\frac{u_{j}}{v_{j}}\alpha_{j}^{n-k-(\ell+1)}&\cdots\\
	\cdots&\frac{u_{j}}{v_{j}}\left(\alpha_{j}^{n-k-\ell}-(-1)^{n-1}P\varOmega_{\ell+1}\right)&\cdots\\ 
	\vdots&\vdots&\vdots\\
	\cdots&\frac{u_{j}}{v_{j}}\left(\alpha_{j}^{n-k}-(-1)^{n-1}P\varOmega_{1}\right)&\cdots
\end{pmatrix}_{(n-k)\times n}.$$
 are $\mathbb{F}_{q}$-linearly independent, it means that $\rank(\boldsymbol{H})=n-k$, and so $$\rank(\boldsymbol{H}_{n-k,*})\geq \rank(\boldsymbol{H})=n-k.$$ Note that $\rank(\boldsymbol{H}_{n-k,*})\leq n-k$, and then
$$\rank(\boldsymbol{H}_{n-k,*})=n-k.$$

Next, we prove $\boldsymbol{G}_{k,*}\boldsymbol{H}_{n-k,*}^{T}=\boldsymbol{0}$. 
For convenience, we denote $\boldsymbol{g}_{i}$ and $\boldsymbol{h}_{j}$ be the $(i+1)$-th $(0\leq i\leq k-1)$ row and the $(j+1)$-th $(0\leq j\leq n-k-1)$ row of $\boldsymbol{G}_{n-k,*}$ and $\boldsymbol{H}_{n-k,*}$, respectively. Then we have the following three cases.

\textbf{Case 1.} For $0\leq i\leq k-1$ and $0\leq j\leq n-k-(\ell+2)$,  we have
$$\boldsymbol{g}_{i}\boldsymbol{h}_{j}^{T}=\begin{cases}
	\sum\limits_{s=1}^{n}u_{s}\left(\alpha_{s}^{j}+\sum\limits_{t=0}^{\ell}\eta_{t}\alpha_{s}^{k+t+j}\right),&\text{if}\ i=0,0\leq j\leq n-k-(\ell+2);\\
	\sum\limits_{s=1}^{n}u_{s}\alpha_{s}^{i+j},&\text{if}\ 1 \leq i\leq k-1,0\leq j\leq n-k-(\ell+2).\\
\end{cases}$$
Note that
$$j\leq n-k-(\ell+2)\leq n-4<n-2,$$
$$k+t+j\leq k+\ell+n-k-(\ell+2)=n-2$$
and 
$$i+j\leq (k-1)+n-k-(\ell+2)=n-\ell-3\leq n-3<n-2,$$
thus, by Lemma \ref{usapowersum}, for $0\leq i\leq k-1$ and $0\leq j\leq n-k-(\ell+2)$,  we have  $\boldsymbol{g}_{i}\boldsymbol{h}_{j}^{T}=0.$ 

\textbf{Case 2.} For $1 \leq i\leq k-1$ and $n-k-(\ell+1)\leq j\leq n-k-1$, we have
$$
\boldsymbol{g}_{i}\boldsymbol{h}_{j}^{T}=\sum\limits_{s=1}^{n}u_{s}\alpha_{s}^{i+j}-\sum\limits_{s=1}^{n}(-1)^{n-1}u_{s}P_{s}\alpha_{s}^{i}\varOmega_{n-k-j}=\sum\limits_{s=1}^{n}u_{s}\alpha_{s}^{i+j}-(-1)^{n-1}P\varOmega_{n-k-j}\sum\limits_{s=1}^{n}u_{s}\alpha_{s}^{i-1}.$$
Note that
$$i+j\leq (k-1)+n-k-1=n-2$$
and
$$i-1\leq k-2\leq n-\ell-3\leq n-3<n-2,$$ 
thus, by Lemma \ref{usapowersum}, for $1 \leq i\leq k-1$ and $n-k-(\ell+1)\leq j\leq n-k-1$, we have $\boldsymbol{g}_{i}\boldsymbol{h}_{j}^{T}=0.$  

\textbf{Case 3.} For $i=0$ and $n-k-(\ell+1)\leq j\leq n-k-1$, for the convenience, we denote $j=n-k-r (1\leq r\leq \ell+1),$ and then we have
$$\begin{aligned}
	&\boldsymbol{g}_{i}\boldsymbol{h}_{j}^{T}\\
	=&\sum\limits_{s=1}^{n}u_{s}\left(1+\sum\limits_{t=0}^{\ell}\eta_{t}\alpha_{s}^{k+t}\right)\left(\alpha_{s}^{n-k-r}-(-1)^{n-1}P_{s}\varOmega_{r}\right)\\
	=&\sum\limits_{s=1}^{n}u_{s}\left(\alpha_{s}^{n-k-r}+\sum\limits_{t=0}^{\ell}\eta_{t}\alpha_{s}^{n+t-r}\right) -\sum\limits_{s=1}^{n}u_{s}\left(1+\sum\limits_{t=0}^{\ell}\eta_{t}\alpha_{s}^{k+t}\right)(-1)^{n-1}P_{s}\varOmega_{r}\\
	=&\sum\limits_{s=1}^{n}u_{s}\alpha_{s}^{n-k-r}+\sum\limits_{s=1}^{n}u_{s}\sum\limits_{t=0}^{\ell}\eta_{t}\alpha_{s}^{n+t-r}-\sum\limits_{s=1}^{n}u_{s}(-1)^{n-1}P_{s}\varOmega_{r}-\sum\limits_{s=1}^{n}u_{s}\sum\limits_{t=0}^{\ell}\eta_{t}\alpha_{s}^{k+t}(-1)^{n-1}P_{s}\varOmega_{r}\\
	=&\sum\limits_{s=1}^{n}u_{s}\alpha_{s}^{n-k-r}+\sum\limits_{t=0}^{\ell}\eta_{t}\sum\limits_{s=1}^{n}u_{s}\alpha_{s}^{n+t-r}-\varOmega_{r}\sum\limits_{s=1}^{n}(-1)^{n-1}u_{s}P_{s}-(-1)^{n-1}\varOmega_{r}\sum\limits_{t=0}^{\ell}\eta_{t}\sum\limits_{s=1}^{n}u_{s}\alpha_{s}^{k+t}P_{s}.
\end{aligned}
$$
Note that $n-k-r\leq n-3<n-2,$ thus by Lemma \ref{usapowersum}, $\sum\limits_{s=1}^{n}u_{s}\alpha_{s}^{n-k-r}=0$. And by Lemma \ref{systemequation}, we have $\sum\limits_{s=1}^{n}(-1)^{n-1}u_{s}P_{s}=1.$ Furthermore, combined $P=\alpha_{s}P_{s}$ and $$k+t-1\leq n-k-(\ell+1)+\ell-1=n-k\leq n-2,$$ then by Lemma  \ref{usapowersum}, for $i=0$ and $n-k-(\ell+1)\leq j\leq n-k-1$, we have
$$\begin{aligned}
	\boldsymbol{g}_{i}\boldsymbol{h}_{j}^{T}=&\sum\limits_{t=0}^{\ell}\eta_{t}\sum\limits_{s=1}^{n}u_{s}\alpha_{s}^{n+t-r}-\varOmega_{r}-(-1)^{n-1}P\varOmega_{r}\sum\limits_{t=0}^{\ell}\eta_{t}\sum\limits_{s=1}^{n}u_{s}\alpha_{s}^{k+t-1}\\
	=&\sum\limits_{t=0}^{\ell}\eta_{t}S_{t+1-r}\left(\alpha_{1},\ldots,\alpha_{n}\right)-\varOmega_{r}=0.
\end{aligned}$$ 

Now by the above three cases,  we can get $\boldsymbol{G}_{k,*}\boldsymbol{H}_{n-k,*}^{T}=\boldsymbol{0}$.
 
From the above discussions, we complete the proof of Theorem $\ref{*LPTGRSparitymatrix}$. 
$\hfill\Box$

\begin{remark}
By taking $\boldsymbol{\eta}=\eta_{0}\in\mathbb{F}_{q}^{*}$, $(\eta_{0},\eta_{1})\in\mathbb{F}_{q}^{2}\backslash\left\{\boldsymbol{0}\right\}$ and $(\eta_{0},\eta_{1},\eta_{2})\in\mathbb{F}_{q}^{3}\backslash\left\{\boldsymbol{0}\right\}$ with $\eta_{1}=0$ in Theorem $\ref{*LPTGRSparitymatrix}$, the corresponding results are just Theorem 1 in \cite{A19}, Lemmas 2.8-2.9 in \cite{A21}, respectively.
\end{remark}

\section{Four classes of LCD codes}
In this section, 
let $n\mid (q-1)$, $\lambda\in\mathbb{F}_{q}^{*}$ with $\mathrm{ord}(\lambda)\mid \frac{q-1}{n}$, and by Lemma \ref{mxroot}, we can set $$m(x)=x^n-\lambda=\prod\limits_{i=1}^{n}(x-\alpha_{i})\in\mathbb{F}_{q}[x]\left( \alpha_{i}\in\mathbb{F}_{q}^{*}\right).$$
Furthermore, by taking the different values of $v_i$ and the dimension $k$ in the generator matrix of $(*)$-$(\mathcal{L},\mathcal{P})$-$\mathrm{TGRS}$ codes, we construct four classes of LCD codes.

\subsection{The main results}
In this subsection, we construct four classes of LCD codes, namely, we prove the following theorems. 
\begin{theorem}\label{LCD1}
	Let $2\leq k\leq \frac{n-2\ell-1}{2}$, $v_{i}\in\left\{-1,1\right\}$ for $k\leq i\leq n$ and  $v_{i}\in\mathbb{F}_{q}\backslash\left\{-1,0,1\right\}$ for $1\leq i\leq k-1$. Then the $(*)$-$(\mathcal{L},\mathcal{P})$-$\mathrm{TGRS}$ code is LCD.
\end{theorem}

\begin{theorem}\label{LCD2}
	Let $2\leq k=\frac{n-\ell-r}{2}(0\leq r\leq \ell)$, $v_{i}\in\left\{-1,1\right\}$ for $k\leq i\leq n$, $v_{i}\in\mathbb{F}_{q}\backslash\left\{-1,0,1\right\}$ for $1\leq i\leq k-1$ and  $1+\sum\limits_{t=r}^{\ell}\eta_{t}\varPhi_{\ell+1+r-t}\neq 0$ with $\varPhi_{i}=(-1)^{n-1}P\varOmega_{i}$. Then the $(*)$-$(\mathcal{L},\mathcal{P})$-$\mathrm{TGRS}$ code  is LCD.
\end{theorem}

\begin{theorem}\label{LCD3}
	Let $2\leq k\leq \frac{n-2\ell-1}{2}$,  $v_{i}\in\left\{-1,1\right\}$ for $1\leq i\leq n-k+1$,   $v_{i}\in\mathbb{F}_{q}\backslash\left\{-1,0,1\right\}$ for $n-k+2\leq i\leq n$. Then the $(*)$-$(\mathcal{L},\mathcal{P})$-$\mathrm{TGRS}$ code  is LCD.
\end{theorem}

\begin{theorem}\label{LCD4}
	Let $2\leq k=\frac{n-\ell-r}{2}(0\leq r\leq \ell)$, $v_{i}\in\left\{-1,1\right\}$ for $1\leq i\leq n-k+1$,   $v_{i}\in\mathbb{F}_{q}\backslash\left\{-1,0,1\right\}$ for $n-k+2\leq i\leq n$ and $1+\sum\limits_{t=r}^{\ell}\eta_{t}\varPhi_{\ell+1+r-t}\neq 0$ with  $\varPhi_{i}=(-1)^{n-1}P\varOmega_{i}(1\leq i\leq (\ell+1))$.  Then the $(*)$-$(\mathcal{L},\mathcal{P})$-$\mathrm{TGRS}$ code is LCD.
\end{theorem}
\subsection{The proofs of main results}
In this subsection, we give the proofs for our main results.
\subsubsection{The proof of Theorem 4.1}
Note that $\boldsymbol{G}_{k,*}$ and $\boldsymbol{H}_{n-k,*}$ given by $(\ref{*LPTGRSgeneratormatrix})$ and $(\ref{*LPTGRSparity-checkmatrix})$ are the generator matrix and the parity-check matrix of $(*)$-$(\mathcal{L},\mathcal{P})$-$\mathrm{TGRS}$ codes, respectively. Now by Lemma \ref{LCDequivalent}, to prove that the $(*)$-$(\mathcal{L},\mathcal{P})$-$\mathrm{TGRS}$ code is LCD, it's enough to prove that  $\rank\begin{pmatrix}
	\boldsymbol{G}_{k,*}\\
	\boldsymbol{H}_{n-k,*}
\end{pmatrix}=n$. Note that  $v_i\in\mathbb{F}_{q}^{*}(1\leq i\leq n)$, then the diagonal matrix $\mathrm{diag}\left\{v_1,\ldots,v_n \right\}$ is nonsingular over $\mathbb{F}_q$, and so it's sufficient to prove that
		$$\rank\left (\boldsymbol{D}\right)=\rank\left(\begin{pmatrix}
			v_{1}& & &\\
			&v_{2}& &\\
			& &\ddots&\\
			& & &v_{n}\\
		\end{pmatrix}\begin{pmatrix}
			\boldsymbol{G}_{k,*}^{T}&\boldsymbol{H}_{n-k,*}^{T}
		\end{pmatrix} \right)=n.$$
For convenience, we denote  $\boldsymbol{D}\triangleq\begin{pmatrix}
			\boldsymbol{A}&\boldsymbol{B}
		\end{pmatrix}$, where
		\begin{equation}\label{LCD1matrixA} 
			\boldsymbol{A}=\begin{pmatrix}
				v_{1}^{2}\left(1+\sum\limits_{t=0}^{\ell}\eta_{t}\alpha_{1}^{k+t}\right)&v_{1}^{2}\alpha_{1}&v_{1}^{2}\alpha_{1}^{2}&\cdots&v_{1}^{2}\alpha_{1}^{k-1}\\
				v_{2}^{2}\left(1+\sum\limits_{t=0}^{\ell}\eta_{t}\alpha_{2}^{k+t}\right)&v_{2}^{2}\alpha_{2}&v_{2}^{2}\alpha_{2}^{2}&\cdots&v_{2}^{2}\alpha_{2}^{k-1}\\
				\vdots&\vdots&\vdots&\vdots\\
				v_{n}^{2}\left(1+\sum\limits_{t=0}^{\ell}\eta_{t}\alpha_{n}^{k+t}\right)&v_{n}^{2}\alpha_{n}&v_{n}^{2}\alpha_{n}^{2}&\cdots&v_{n}^{2}\alpha_{n}^{k-1}
			\end{pmatrix}
		\end{equation}
and
		\begin{equation}\label{LCD1matrixB}\small
			\boldsymbol{B}=\begin{pmatrix}
				u_{1}\alpha_{1}^{0}&\cdots&u_{1}\alpha_{1}^{n-k-(\ell +2)}&u_{1}\left(\alpha_{1}^{n-k-(\ell +1)}-\varDelta_{(\ell+1)1}\right)&\cdots&u_{1}\left(\alpha_{1}^{n-k-1}-\varDelta_{11}\right)\\ 
				u_{2}\alpha_{2}^{0}&\cdots&u_{2}\alpha_{2}^{n-k-(\ell +2)}&u_{2}\left(\alpha_{2}^{n-k-(\ell +1)}-\varDelta_{(\ell+1)2}\right)&\cdots&u_{2}\left(\alpha_{2}^{n-k-1}-\varDelta_{12}\right)\\ 
				\vdots&\ddots&\vdots&\vdots&\ddots&\vdots\\
				u_{n}\alpha_{n}^{0}&\cdots&u_{n}\alpha_{n}^{n-k-(\ell +2)}&u_{n}\left(\alpha_{n}^{n-k-(\ell +1)}-\varDelta_{(\ell+1)n}\right)&\cdots&u_{n}\left(\alpha_{n}^{n-k-1}-\varDelta_{1n}\right)\\ 
			\end{pmatrix}
		\end{equation}
		with $$\varDelta_{ij}=(-1)^{n-1}P_{j}\varOmega_{i}=(-1)^{n-1}\frac{P}{\alpha_{j}}\varOmega_{i}.$$
		Note that $m(x)=x^n-\lambda=\prod\limits_{i=1}^{n}(x-\alpha_{i})\in\mathbb{F}_{q}[x]$, thus we have
		$$\alpha_{i}^{n}=\lambda$$ 
		and
		$$m^{\prime}(x)=nx^{n-1}=\sum\limits_{i=1}^{n}\prod\limits_{j=1,j\neq i}^{n}(x-\alpha_{j}),$$ then 
		\begin{equation}\label{uilambda}
		u_{i}=\frac{1}{m^{\prime}(\alpha_{i})}=\frac{1}{n\alpha_{i}^{n-1}}=\frac{1}{n\lambda}\alpha_{i}.
		\end{equation}
		Thus by $v_{i}\in\mathbb{F}_{q}\backslash\left\{-1,0,1\right\}$ for $1\leq i\leq k-1$ and  $v_{i}\in\left\{-1,1\right\}$ for $k\leq i\leq n$, then it's enough to prove that
		$$\rank\begin{pmatrix}
			\boldsymbol{A}_{1}^{(1)}&\boldsymbol{B}_{1}^{(1)}
		\end{pmatrix}=n,$$
		where
		\begin{equation}\label{LCD1matrixA1(1)} 
			\boldsymbol{A}_{1}^{(1)}=\begin{pmatrix}
				v_{1}^{2}\left(1+\sum\limits_{t=0}^{\ell}\eta_{t}\alpha_{1}^{k+t}\right)&v_{1}^{2}\alpha_{1}&v_{1}^{2}\alpha_{1}^{2}&\cdots&v_{1}^{2}\alpha_{1}^{k-1}\\
				\vdots&\vdots&\vdots&\vdots\\
				v_{k-1}^{2}\left(1+\sum\limits_{t=0}^{\ell}\eta_{t}\alpha_{k-1}^{k+t}\right)&v_{k-1}^{2}\alpha_{k-1}&v_{k-1}^{2}\alpha_{k-1}^{2}&\cdots&v_{k-1}^{2}\alpha_{k-1}^{k-1}\\
				1+\sum\limits_{t=0}^{\ell}\eta_{t}\alpha_{k}^{k+t}&\alpha_{k}&\alpha_{k}^{2}&\cdots&\alpha_{k}^{k-1}\\
				\vdots&\vdots&\vdots&\vdots\\
				1+\sum\limits_{t=0}^{\ell}\eta_{t}\alpha_{n}^{k+t}&\alpha_{n}&\alpha_{n}^{2}&\cdots&\alpha_{n}^{k-1}\\
			\end{pmatrix},
		\end{equation}
and
		\begin{equation}\label{LCD1matrixB1(1)}
			\boldsymbol{B}_{1}^{(1)}=\begin{pmatrix}
				\frac{\alpha_{1}}{n\lambda}&\frac{\alpha_{1}^{2}}{n\lambda}&\cdots&\frac{\alpha_{1}^{n-k-(\ell +1)}}{n\lambda}&\frac{\alpha_{1}^{n-k-\ell}-\varDelta_{(\ell+1)1}\alpha_{1}}{n\lambda}&\cdots&\frac{\alpha_{1}^{n-k}-\varDelta_{11}\alpha_{1}}{n\lambda}\\ 
				\frac{\alpha_{2}}{n\lambda}&\frac{\alpha_{2}^{2}}{n\lambda}&\cdots&\frac{\alpha_{2}^{n-k-(\ell +1)}}{n\lambda}&\frac{\alpha_{2}^{n-k-\ell}-\varDelta_{(\ell+1)2}\alpha_{2}}{n\lambda}&\cdots&\frac{\alpha_{2}^{n-k}-\varDelta_{12}\alpha_{2}}{n\lambda}\\
				\vdots&\vdots&\ddots&\vdots&\vdots&\ddots&\vdots\\
				\frac{\alpha_{n}}{n\lambda}&\frac{\alpha_{n}^{2}}{n\lambda}&\cdots&\frac{\alpha_{n}^{n-k-(\ell +1)}}{n\lambda}&\frac{\alpha_{n}^{n-k-\ell}-\varDelta_{(\ell+1)n}\alpha_{n}}{n\lambda}&\cdots&\frac{\alpha_{n}^{n-k}-\varDelta_{1n }\alpha_{n}}{n\lambda}\\
			\end{pmatrix}
		\end{equation}
with $\varDelta_{ij}=(-1)^{n-1}P_{j}\varOmega_{i}=(-1)^{n-1}\frac{P}{\alpha_{j}}\varOmega_{i}.$
		
		Now, for the matrix $\boldsymbol{B}_{1}^{(1)}$, by multiplying $n\lambda\in\mathbb{F}_{q}^{*}$ to the each row, we can obtain the matrix  \begin{equation}\label{LCD1matrixB1(2)}
			\boldsymbol{B}_{1}^{(2)}=\begin{pmatrix}
				\alpha_{1}&\alpha_{1}^{2}&\cdots&\alpha_{1}^{n-k-(\ell +1)}&\alpha_{1}^{n-k-\ell}-\varPhi_{\ell+1}&\cdots&\alpha_{1}^{n-k}-\varPhi_{1}\\ 
				\alpha_{2}&\alpha_{2}^{2}&\cdots&\alpha_{2}^{n-k-(\ell +1)}&\alpha_{2}^{n-k-\ell}-\varPhi_{\ell+1}&\cdots&\alpha_{2}^{n-k}-\varPhi_{1}\\
				\vdots&\vdots&\ddots&\vdots&\vdots&\ddots&\vdots\\
				\alpha_{n}&\alpha_{n}^{2}&\cdots&\alpha_{n}^{n-k-(\ell +1)}&\alpha_{n}^{n-k-\ell}-\varPhi_{\ell+1}&\cdots&\alpha_{n}^{n-k}-\varPhi_{1}\\
			\end{pmatrix},
		\end{equation}
		where $\varPhi_i=(-1)^{n-1}P\varOmega_{i}.$ And
$$\rank\begin{pmatrix}
	\boldsymbol{A}_{1}^{(1)}&\boldsymbol{B}_{1}^{(1)}
\end{pmatrix}=\rank\begin{pmatrix}
	\boldsymbol{A}_{1}^{(1)}&\boldsymbol{B}_{1}^{(2)}
\end{pmatrix}.$$
		
		Note that $ 2\leq k\leq\frac{n-2\ell-1}{2}$, it means that $$n-k-(\ell+1)\geq k+\ell\geq k>k-1.$$
Therefore, for  the matrix $\begin{pmatrix}
	\boldsymbol{A}_{1}^{(1)}&\boldsymbol{B}_{1}^{(2)}
\end{pmatrix}$, we can carry out the following elementary transformations. For convenience, we denote the vector $\boldsymbol{a}_{i,1}^{(1)}$ and $\boldsymbol{b}_{j,1}^{(2)}$ be the $(i+1)$-th $(0\leq i\leq k-1)$ column and the $j$-th $(1\leq j\leq n-k)$ column of $\boldsymbol{A}_{1}^{(1)}$ and $\boldsymbol{B}_{1}^{(2)}$, respectively.
		
\textbf{Step 1.} Replace $\boldsymbol{a}_{i+1,1}^{(1)}$ with $\boldsymbol{a}_{i+1,1}^{(1)}-\boldsymbol{b}_{i,1}^{(2)}$ for each $i=1,\ldots,k-1$, we can obtain the matrix $\begin{pmatrix}
	\boldsymbol{A}_{1}^{(2)}&\boldsymbol{B}_{1}^{(3)}
\end{pmatrix}$ and denote the vector $\boldsymbol{a}_{i,1}^{(2)}$ and $\boldsymbol{b}_{j,1}^{(3)}$ be the $(i+1)$-th $(0\leq i\leq k-1)$ column and the $j$-th $(1\leq j\leq n-k)$ column of $\boldsymbol{A}_{1}^{(2)}$ and $\boldsymbol{B}_{1}^{(3)}$, respectively; 
		
\textbf{Step 2.} Replace $\boldsymbol{a}_{1,1}^{(2)}$ with $\boldsymbol{a}_{1,1}^{(2)}-\sum\limits_{t=0}^{\ell}\eta_t\boldsymbol{b}_{k+t,1}^{(3)}$, we can obtain the matrix $\begin{pmatrix}
	\boldsymbol{A}_{1}^{(3)}&\boldsymbol{B}_{1}^{(4)}
\end{pmatrix}$; 
 
\textbf{Step 3.} put the first column of the matrix $\boldsymbol{A}_{1}^{(3)}$ to the back of the last column of the matrix $\boldsymbol{A}_{1}^{(3)}$, we can obtain the matrix $\begin{pmatrix}
			\boldsymbol{A}_{1}^{(4)}&\boldsymbol{B}_{1}^{(5)}
		\end{pmatrix}$, 
where
		\begin{equation}\label{LCD1matrixA1(2)}
			\boldsymbol{A}_{1}^{(4)}=\begin{pmatrix}
				\left(v_{1}^{2}-1\right)\alpha_{1}&\cdots&\left(v_{1}^{2}-1\right)\alpha_{1}^{k-1}\\
				\vdots&\ddots&\vdots\\
				\left(v_{k-1}^{2}-1\right)\alpha_{k-1}&\cdots&\left(v_{k-1}^{2}-1\right)\alpha_{k-1}^{k-1}\\ 
				0&\cdots&0\\
				
				\vdots&\ddots&\vdots\\
				
				0&\cdots&0\\
			\end{pmatrix},
		\end{equation}
and
		$$
		\boldsymbol{B}_{1}^{(5)}=\begin{pmatrix}
			\varTheta_{1}&\alpha_{1}&\cdots&\alpha_{1}^{n-k-(\ell+1)}&\alpha_{1}^{n-k-\ell}-\varPhi_{\ell+1}&\cdots&\alpha_{1}^{n-k}-\varPhi_{1}\\ 
			\vdots&\vdots&\ddots&
			\vdots&\vdots&\ddots&\vdots\\
			\varTheta_{k-1}&\alpha_{k-1}&\cdots&\alpha_{k-1}^{n-k-(\ell+1)}&\alpha_{k-1}^{n-k-\ell}-\varPhi_{\ell+1}&\cdots&\alpha_{k-1}^{n-k}-\varPhi_{1}\\
			1&\alpha_{k}&\cdots&\alpha_{k}^{n-k-(\ell+1)}&\alpha_{k}^{n-k-\ell}-\varPhi_{\ell+1}&\cdots&\alpha_{k}^{n-k}-\varPhi_{1}\\
			\vdots&\vdots&\ddots&
			\vdots&\vdots&\ddots&\vdots\\
			1&\alpha_{n}&\cdots&\alpha_{n}^{n-k-(\ell+1)}&\alpha_{n}^{n-k-\ell}-\varPhi_{\ell+1}&\cdots&\alpha_{n}^{n-k}-\varPhi_{1}\\
		\end{pmatrix}
		$$
with $\varTheta_{i}=\left(v_{i}^{2}-1\right)\left(1+\sum\limits_{t=0}^{\ell}\eta_{t}\alpha_{i}^{k+t}\right)+1.$ 
		
		Note that for any matrix $\boldsymbol{M}$, the rank of $\boldsymbol{M}$ cannot be changed under the elementary transformations  , and so it's enough to prove that 
		$$\rank\begin{pmatrix}
			\boldsymbol{A}_{1}^{(4)}&\boldsymbol{B}_{1}^{(5)}
		\end{pmatrix}=n.$$  
i.e., $\det\begin{pmatrix}
	\boldsymbol{A}_{1}^{(4)}&\boldsymbol{B}_{1}^{(5)}
\end{pmatrix}\neq 0$. For convenience, we denote $\begin{pmatrix}
			\boldsymbol{A}_{1}^{(4)}&\boldsymbol{B}_{1}^{(5)}
		\end{pmatrix}\triangleq\begin{pmatrix}
			\boldsymbol{K}_{1}&*\\
			\boldsymbol{0}&\boldsymbol{L}_{1}
		\end{pmatrix}$, where
		\begin{equation}\label{LCD1k1}
			\boldsymbol{K}_{1}=\begin{pmatrix}
				\left(v_{1}^{2}-1\right)\alpha_{1}&\cdots&\left(v_{1}^{2}-1\right)\alpha_{1}^{k-1}\\
				\vdots&\ddots&\vdots\\
				\left(v_{k-1}^{2}-1\right)\alpha_{k-1}&\cdots&\left(v_{k-1}^{2}-1\right)\alpha_{k-1}^{k-1}
			\end{pmatrix}_{(k-1)\times (k-1)}
		\end{equation}
		and 
		$$
		\boldsymbol{L}_{1}=\begin{pmatrix}
			1&\alpha_{k}&\cdots&\alpha_{k}^{n-k-(\ell+1)}&\alpha_{k}^{n-k-\ell}-\varPhi_{\ell+1}&\cdots&\alpha_{k}^{n-k}-\varPhi_{1}\\
			\vdots&\vdots&\ddots&
			\vdots&\vdots&\ddots&\vdots\\
			1&\alpha_{n}&\cdots&\alpha_{n}^{n-k-(\ell+1)}&\alpha_{n}^{n-k-\ell}-\varPhi_{\ell+1}&\cdots&\alpha_{n}^{n-k}-\varPhi_{1}\\
		\end{pmatrix}_{(n-k+1)\times (n-k+1)}.
		$$ 
		Note that
		\begin{equation}\label{LCD1detK1}
			\det(\boldsymbol{K}_{1})=
			\prod\limits_{i=1}^{k-1}\left(v_{i}^{2}-1\right)\cdot\prod\limits_{i=1}^{k-1}\alpha_{i}\cdot\prod\limits_{1\leq a<b\leq k-1}(\alpha_{b}-\alpha_{a})\neq 0
		\end{equation}
		and
		$$\det(\boldsymbol{L}_{1})=\det\begin{pmatrix}
			1&\alpha_{k}&\cdots&\alpha_{k}^{n-k-(\ell+1)}&\alpha_{k}^{n-k-\ell}&\cdots&\alpha_{k}^{n-k}\\
			\vdots&\vdots&\ddots&
			\vdots&\vdots&\ddots&\vdots\\
			1&\alpha_{n}&\cdots&\alpha_{n}^{n-k-(\ell+1)}&\alpha_{n}^{n-k-\ell}&\cdots&\alpha_{n}^{n-k}\\
		\end{pmatrix}=\prod\limits_{k\leq a<b\leq n}(\alpha_{b}-\alpha_{a})\neq 0.
		$$
		Hence $$ \det\begin{pmatrix}
			\boldsymbol{A}_{1}^{(4)}&\boldsymbol{B}_{1}^{(5)}
		\end{pmatrix}=\det(\boldsymbol{K}_{1})\det(\boldsymbol{L}_{1})\neq 0,$$
		it means $\rank\begin{pmatrix}
			\boldsymbol{A}_{1}^{(4)}&\boldsymbol{B}_{1}^{(5)}
		\end{pmatrix}=n$, thus 
		$$\rank\begin{pmatrix}
			\boldsymbol{G}_{k,*}^{T}&
			\boldsymbol{H}_{n-k,*}^{T}
		\end{pmatrix}=\rank\begin{pmatrix}
			\boldsymbol{G}_{k,*}\\
			\boldsymbol{H}_{n-k,*}
		\end{pmatrix}=\rank\left(\boldsymbol{D}\right)=\rank\begin{pmatrix}
			\boldsymbol{A}_{1}^{(4)}&\boldsymbol{B}_{1}^{(5)}
		\end{pmatrix}=n,$$
which means that 
		the $(*)$-$(\mathcal{L},\mathcal{P})$-$\mathrm{TGRS}$ code is LCD. 
		
		From the above, we complete the proof of Theorem $\ref{LCD1}$. 

\subsubsection{The proof of Theorem 4.2}
In the similar proof as that of Theorem \ref{LCD1}, it's sufficient to prove that
$$\rank\begin{pmatrix}
	\boldsymbol{A}_{1}^{(1)}&\boldsymbol{B}_{1}^{(2)}
\end{pmatrix}=n,$$
where $\boldsymbol{A}_{1}^{(1)}$ and $\boldsymbol{B}_{1}^{(2)}$ are given by $(\ref{LCD1matrixA1(1)})$ and  $(\ref{LCD1matrixB1(2)})$, respectively. 

Note that $n=2k+\ell+r(0\leq r\leq \ell)$, thus we can get
$$n-k-(\ell+1)=k+(r-1)\geq k-1,$$
$$n-k-\ell=k+r$$ 
and
$$n-k=k+\ell+r\geq k+\ell.$$
Then we can carry out the following elementary transformations:

\textbf{Step 1.}  Replace $\boldsymbol{a}_{i+1,1}^{(1)}$ with $\boldsymbol{a}_{i+1,1}^{(1)}-\boldsymbol{b}_{i,1}^{(2)}$ for each $i=1,\ldots,k-1$, we can obtain the matrix $\begin{pmatrix}
	\boldsymbol{A}_{1}^{(2)}&\boldsymbol{B}_{1}^{(3)}
\end{pmatrix}$ and denote the vector $\boldsymbol{a}_{i,1}^{(2)}$ and $\boldsymbol{b}_{j,1}^{(3)}$ be the $(i+1)$-th $(0\leq i\leq k-1)$ column and the $j$-th $(1\leq j\leq n-k)$ column of $\boldsymbol{A}_{1}^{(2)}$ and $\boldsymbol{B}_{1}^{(3)}$, respectively; 

\textbf{Step 2.}  Replace $\boldsymbol{a}_{1,1}^{(2)}$ with $\boldsymbol{a}_{1,1}^{(2)}-\sum\limits_{t=0}^{r-1}\eta_{t}\boldsymbol{b}_{k+t,1}^{(3)}$, we can obtain the matrix $\begin{pmatrix}
	\boldsymbol{A}_{1}^{(5)}&\boldsymbol{B}_{1}^{(6)}
\end{pmatrix}$; 

\textbf{Step 3.}  put the first column of the matrix $\boldsymbol{A}_{1}^{(5)}$ to the back of the last column of the matrix $\boldsymbol{A}_{1}^{(5)}$, we can obtain the matrix $\begin{pmatrix}
	\boldsymbol{A}_{1}^{(4)}&\boldsymbol{B}_{1}^{(7)}
\end{pmatrix},$ where $\boldsymbol{A}_{1}^{(4)}$ is given by $(\ref{LCD1matrixA1(2)})$ and
$$
\boldsymbol{B}_{1}^{(7)}=\begin{pmatrix}
	\varXi_{1}&\alpha_{1}&\cdots&\alpha_{1}^{n-k-(\ell+1)}&\alpha_{1}^{n-k-\ell}-\varPhi_{\ell+1}&\cdots&\alpha_{1}^{n-k}-\varPhi_{1}\\ 
	\vdots&\vdots&\ddots&
	\vdots&\vdots&\ddots&\vdots\\
	\varXi_{k-1}&\alpha_{k-1}&\cdots&\alpha_{k-1}^{n-k-(\ell+1)}&\alpha_{k-1}^{n-k-\ell}-\varPhi_{\ell+1}&\cdots&\alpha_{k-1}^{n-k}-\varPhi_{1}\\
	1+\sum\limits_{t=r}^{\ell}\eta_{t}\varPhi_{\ell+1+r-t}&\alpha_{k}&\cdots&\alpha_{k}^{n-k-(\ell+1)}&\alpha_{k}^{n-k-\ell}-\varPhi_{\ell+1}&\cdots&\alpha_{k}^{n-k}-\varPhi_{1}\\
	\vdots&\vdots&\ddots&
	\vdots&\vdots&\ddots&\vdots\\
	1+\sum\limits_{t=r}^{\ell}\eta_{t}\varPhi_{\ell+1+r-t}&\alpha_{n}&\cdots&\alpha_{n}^{n-k-(\ell+1)}&\alpha_{n}^{n-k-\ell}-\varPhi_{\ell+1}&\cdots&\alpha_{n}^{n-k}-\varPhi_{1}\\
\end{pmatrix}
$$ 
with $\varXi_{i}=\left(v_{i}^{2}-1\right)\left(1+\sum\limits_{t=0}^{\ell}\eta_{t}\alpha_{i}^{k+t}\right)+1+\sum\limits_{t=r}^{\ell}\eta_{t}\varPhi_{\ell+1+r-t}.$

Note that for any matrix $\boldsymbol{M}$, the rank of $\boldsymbol{M}$ cannot be changed under the elementary transformations  , and so it's enough to prove that 
$$\rank\begin{pmatrix}
	\boldsymbol{A}_{1}^{(4)}&\boldsymbol{B}_{1}^{(7)}
\end{pmatrix}=n.$$ 
i.e., $\det\begin{pmatrix}
	\boldsymbol{A}_{1}^{(4)}&\boldsymbol{B}_{1}^{(7)}
\end{pmatrix}\neq 0$. For convenience, we denote $\begin{pmatrix}
	\boldsymbol{A}_{1}^{(4)}&\boldsymbol{B}_{1}^{(7)}
\end{pmatrix}=\begin{pmatrix}
	\boldsymbol{K}_{1}&*\\
	\boldsymbol{0}&\boldsymbol{L}_{2}
\end{pmatrix},$
where $\boldsymbol{K}_{1}$ is given by $(\ref{LCD1k1})$ and 
$$
\boldsymbol{L}_{2}=\begin{pmatrix}
	1+\sum\limits_{t=r}^{\ell}\eta_{t}\varPhi_{\ell+1+r-t}&\alpha_{k}&\cdots&\alpha_{k}^{n-k-(\ell+1)}&\alpha_{k}^{n-k-\ell}-\varPhi_{\ell+1}&\cdots&\alpha_{k}^{n-k}-\varPhi_{1}\\
	\vdots&\vdots&\ddots&
	\vdots&\vdots&\ddots&\vdots\\
	1+\sum\limits_{t=r}^{\ell}\eta_{t}\varPhi_{\ell+1+r-t}&\alpha_{n}&\cdots&\alpha_{n}^{n-k-(\ell+1)}&\alpha_{n}^{n-k-\ell}-\varPhi_{\ell+1}&\cdots&\alpha_{n}^{n-k}-\varPhi_{1}\\
\end{pmatrix}.
$$
By $(\ref{LCD1detK1})$, we have $\det(\boldsymbol{K}_{1})\neq 0$
and
$$\small\begin{aligned}
	&\det(\boldsymbol{L}_{2})\\
	=&\left(1+\sum\limits_{t=r}^{\ell}\eta_{t}\varPhi_{\ell+1+r-t}\right)\cdot\det\begin{pmatrix}
		1&\alpha_{k}&\cdots&\alpha_{k}^{n-k-(\ell+1)}&\alpha_{k}^{n-k-\ell}-\varPhi_{\ell+1}&\cdots&\alpha_{k}^{n-k}-\varPhi_{1}\\
		\vdots&\vdots&\ddots&
		\vdots&\vdots&\ddots&\vdots\\
		1&\alpha_{n}&\cdots&\alpha_{n}^{n-k-(\ell+1)}&\alpha_{n}^{n-k-\ell}-\varPhi_{\ell+1}&\cdots&\alpha_{n}^{n-k}-\varPhi_{1}\\
	\end{pmatrix}\\
	=&\left(1+\sum\limits_{t=r}^{\ell}\eta_{t}\varPhi_{\ell+1+r-t}\right)\cdot\det\begin{pmatrix}
		1&\alpha_{k}&\cdots&\alpha_{k}^{n-k-(\ell+1)}&\alpha_{k}^{n-k-\ell}&\cdots&\alpha_{k}^{n-k}\\
		\vdots&\vdots&\ddots&
		\vdots&\vdots&\ddots&\vdots\\
		1&\alpha_{n}&\cdots&\alpha_{n}^{n-k-(\ell+1)}&\alpha_{n}^{n-k-\ell}&\cdots&\alpha_{n}^{n-k}\\
	\end{pmatrix}\\
	=&\left(1+\sum\limits_{t=r}^{\ell}\eta_{t}\varPhi_{\ell+1+r-t}\right)\cdot\prod\limits_{k\leq a<b\leq n}(\alpha_{b}-\alpha_{a})\neq 0.
\end{aligned}
$$
Thus by $1+\sum\limits_{t=r}^{\ell}\eta_{t}\varPhi_{\ell+1+r-t}\neq 0$, we have
$$ \det\begin{pmatrix}
	\boldsymbol{A}_{1}^{(4)}&\boldsymbol{B}_{1}^{(7)}
\end{pmatrix}=\det(\boldsymbol{K}_{1})\cdot\det(\boldsymbol{L}_{2})\neq 0,$$
it means $\rank\begin{pmatrix}
	\boldsymbol{A}_{1}^{(4)}&\boldsymbol{B}_{1}^{(7)}
\end{pmatrix}=n$, i.e., 
$$\rank\begin{pmatrix}
	\boldsymbol{G}_{k,*}^{T}&
	\boldsymbol{H}_{n-k,*}^{T}
\end{pmatrix}=\rank\begin{pmatrix}
	\boldsymbol{G}_{k,*}\\
	\boldsymbol{H}_{n-k,*}
\end{pmatrix}=\rank(\boldsymbol{D})=\rank\begin{pmatrix}
	\boldsymbol{A}_{1}^{(4)}&\boldsymbol{B}_{1}^{(7)}
\end{pmatrix}=n,$$
which means that the  
$(*)$-$(\mathcal{L},\mathcal{P})$-$\mathrm{TGRS}$ code is LCD . 

From the above, we complete the proof of Theorem $\ref{LCD2}$.

\subsubsection{The proof of Theorem 4.3}
{\bf Proof}. In the similar proof as that of Theorem \ref{LCD1}, and by $v_{i}\in\left\{-1,1\right\}$ for $1\leq i\leq n-k+1$,  $v_{i}\in\mathbb{F}_{q}\backslash\left\{-1,0,1\right\}$ for $n-k+2\leq i\leq n$ and $ (\ref{uilambda})$, we can obtain the matrix  $\begin{pmatrix}
	\boldsymbol{A}_{2}^{(1)}&\boldsymbol{B}_{1}^{(2)}
\end{pmatrix},$
where 
\begin{equation}\label{LCD3matrixA2(1)}
	\boldsymbol{A}_{2}^{(1)}=\begin{pmatrix}
		1+\sum\limits_{t=0}^{\ell}\eta_{t}\alpha_{1}^{k+t}& \alpha_{1}& \cdots& \alpha_{1}^{k-1}\\
		\vdots&\vdots&\vdots\\
		1+\sum\limits_{t=0}^{\ell}\eta_{t}\alpha_{n-k+1}^{k+t}&\alpha_{n-k+1} &\cdots&\alpha_{n-k+1}^{k-1}\\
		v_{n-k+2}^{2}\left(1+\sum\limits_{t=0}^{\ell}\eta_{t}\alpha_{n-k+2}^{k+t}\right)&v_{n-k+2}^{2}\alpha_{n-k+2}& \cdots&v_{n-k+2}^{2}\alpha_{n-k+2}^{k-1}\\
		\vdots&\vdots&\vdots&\vdots\\
		v_{n}^{2}\left(1+\sum\limits_{t=0}^{\ell}\eta_{t}\alpha_{n}^{k+t}\right)&v_{n}^{2}\alpha_{n} &\cdots&v_{n}^{2}\alpha_{n}^{k-1}\\
	\end{pmatrix}
\end{equation} 
  and $\boldsymbol{B}_{1}^{(2)}$ is given by $(\ref{LCD1matrixB1(2)})$. And then it's enough to prove that
  $$\rank\begin{pmatrix}
  	\boldsymbol{A}_{2}^{(1)}&\boldsymbol{B}_{1}^{(2)}
  \end{pmatrix}=n.$$
 
By $2\leq k\leq \frac{n-2\ell-1}{2}$, we have $$n-k-(\ell+1)\geq k+\ell>k-1.$$ Therefore, for  the matrix $\begin{pmatrix}
	\boldsymbol{A}_{2}^{(1)}&\boldsymbol{B}_{1}^{(2)}
\end{pmatrix}$, we can carry out the following elementary transformations. For convenience, we denote the vector $\boldsymbol{a}_{i,2}^{(1)}$ and $\boldsymbol{b}_{j,1}^{(2)}$ be the $(i+1)$-th $(0\leq i\leq k-1)$ column and the $j$-th $(1\leq j\leq n-k)$ column of $\boldsymbol{A}_{2}^{(1)}$ and $\boldsymbol{B}_{1}^{(2)}$, respectively.

\textbf{Step 1.} Replace $\boldsymbol{a}_{i+1,2}^{(1)}$ with $\boldsymbol{a}_{i+1,2}^{(2)}-\boldsymbol{b}_{i,1}^{(2)}$ for each $i=1,\ldots,k-1$, we can obtain the matrix $\begin{pmatrix}
	\boldsymbol{A}_{2}^{(2)}&\boldsymbol{B}_{1}^{(8)}
\end{pmatrix}$ and denote the vector $\boldsymbol{a}_{i,2}^{(2)}$ and $\boldsymbol{b}_{j,1}^{(8)}$ be the $(i+1)$-th $(0\leq i\leq k-1)$ column and the $j$-th $(1\leq j\leq n-k)$ column of $\boldsymbol{A}_{2}^{(2)}$ and $\boldsymbol{B}_{1}^{(8)}$, respectively; 

\textbf{Step 2.} Replace $\boldsymbol{a}_{1,2}^{(2)}$ with $\boldsymbol{a}_{1,2}^{(2)}-\sum\limits_{t=0}^{\ell}\eta_t\boldsymbol{b}_{k+t,1}^{(8)}$, we can obtain the matrix $\begin{pmatrix}
	\boldsymbol{A}_{2}^{(3)}&\boldsymbol{B}_{1}^{(9)}
\end{pmatrix}$; 

\textbf{Step 3.} put the first column of the matrix $\boldsymbol{A}_{2}^{(3)}$ to the back of the last column of the matrix $\boldsymbol{A}_{2}^{(3)}$, we can obtain the matrix $\begin{pmatrix}
	\boldsymbol{A}_{2}^{(4)}&\boldsymbol{B}_{1}^{(10)}
\end{pmatrix},$
where
\begin{equation}\label{LCD3matrixA2(2)}
	\boldsymbol{A}_{2}^{(4)}=\begin{pmatrix}
		0&\cdots&0\\
		\vdots& &\vdots\\
		0&\cdots&0\\
		\left(v_{n-k+2}^{2}-1\right)\alpha_{n-k+2}&\cdots&\left(v_{n-k+2}^{2}-1\right)\alpha_{n-k+2}^{k-1}\\
		\vdots& &\vdots\\
		\left(v_{n}^{2}-1\right)\alpha_{n}&\cdots&\left(v_{n}^{2}-1\right)\alpha_{n}^{k-1}\\
	\end{pmatrix}
\end{equation}
and
$$
\boldsymbol{B}_{1}^{(10)}=\begin{pmatrix}
	1&\alpha_{1}&\cdots&\alpha_{1}^{n-k-(\ell +1)}&\alpha_{1}^{n-k-\ell}-\varPhi_{\ell+1}&\cdots&\alpha_{1}^{n-k}-\varPhi_{1}\\ 
	\vdots&\vdots&\ddots&\vdots&\vdots&\ddots&\vdots\\
	1&\alpha_{n-k+1}&\cdots&\alpha_{n-k+1}^{n-k-(\ell +1)}&\alpha_{n-k+1}^{n-k-\ell}-\varPhi_{\ell+1}&\cdots&\alpha_{n-k+1}^{n-k}-\varPhi_{1}\\
	\varTheta_{n-k+2}&\alpha_{n-k+2}&\cdots&\alpha_{n-k+2}^{n-k-(\ell +1)}&\alpha_{n-k+2}^{n-k-\ell}-\varPhi_{\ell+1}&\cdots&\alpha_{n-k+2}^{n-k}-\varPhi_{1}\\
	\vdots&\vdots&\ddots&\vdots&\vdots&\ddots&\vdots\\
	\varTheta_{n}&\alpha_{n}&\cdots&\alpha_{n}^{n-k-(\ell +1)}&\alpha_{n}^{n-k-\ell}-\varPhi_{\ell+1}&\cdots&\alpha_{n}^{n-k}-\varPhi_{1}\\
\end{pmatrix}
$$ 
with $\varTheta_{i}=\left(v_{i}^{2}-1\right)\left(1+\sum\limits_{t=0}^{\ell}\eta_{t}\alpha_{i}^{k+t}\right)+1.$ 

Note that for any matrix $\boldsymbol{M}$, the rank of $\boldsymbol{M}$ cannot be changed under the elementary transformations  , and so it's enough to prove that   
$$\rank\begin{pmatrix}
	\boldsymbol{A}_{2}^{(4)}&\boldsymbol{B}_{1}^{(10)}
\end{pmatrix}=n,$$
i.e., $\det\begin{pmatrix}
	\boldsymbol{A}_{2}^{(4)}&\boldsymbol{B}_{1}^{(10)}
\end{pmatrix}\neq 0$. For convenience, we denote $\begin{pmatrix}
	\boldsymbol{A}_{2}^{(4)}&\boldsymbol{B}_{1}^{(10)}
\end{pmatrix}=\begin{pmatrix}
	\boldsymbol{0}&\boldsymbol{L}_{3}\\
	\boldsymbol{K}_{2}&*
\end{pmatrix}$,
where
\begin{equation}\label{LCD3K2}
	\boldsymbol{K}_{2}=\begin{pmatrix}
		\left(v_{n-k+2}^{2}-1\right)\alpha_{n-k+2}&\cdots&\left(v_{n-k+2}^{2}-1\right)\alpha_{n-k+2}^{k-1}\\
		\vdots&\ddots&\vdots\\
		\left(v_{n}^{2}-1\right)\alpha_{n}&\cdots&\left(v_{n}^{2}-1\right)\alpha_{n}^{k-1}\\
	\end{pmatrix}_{(k-1)\times (k-1)}
\end{equation}
and
$$
\boldsymbol{L}_{3}=\begin{pmatrix}
	1&\alpha_{1}& \cdots&\alpha_{1}^{n-k-(\ell +1)}&\alpha_{1}^{n-k-\ell}-\varPhi_{\ell+1}&\cdots&\alpha_{1}^{n-k}-\varPhi_{1}\\ 
	\vdots&\vdots&\ddots&\vdots&\vdots&\ddots&\vdots\\
	1&\alpha_{n-k+1} &\cdots&\alpha_{n-k+1}^{n-k-(\ell +1)}&\alpha_{n-k+1}^{n-k-\ell}-\varPhi_{\ell+1}&\cdots&\alpha_{n-k+1}^{n-k}-\varPhi_{1}\\
\end{pmatrix}_{(n-k+1)\times (n-k+1)}.
$$  
Note that
\begin{equation}\label{K2det}
\det(\boldsymbol{K}_{2})=\prod\limits_{i=n-k+2}^{n}\left(v_{i}^{2}-1\right)\cdot\prod\limits_{i=n-k+2}^{n}\alpha_{i}\cdot\prod\limits_{n-k+2\leq a<b\leq n}(\alpha_{b}-\alpha_{a})\neq 0
\end{equation}
and
$$\det(\boldsymbol{L}_{3})=\det\begin{pmatrix}
		1&\alpha_{1}& \cdots&\alpha_{1}^{n-k-(\ell +1)}&\alpha_{1}^{n-k-\ell}&\cdots&\alpha_{1}^{n-k}\\ 
		\vdots&\vdots&\ddots&\vdots&\vdots&\ddots&\vdots\\
		1&\alpha_{n-k+1} &\cdots&\alpha_{n-k+1}^{n-k-(\ell +1)}&\alpha_{n-k+1}^{n-k-\ell}&\cdots&\alpha_{n-k+1}^{n-k}\\
	\end{pmatrix}=\prod\limits_{1\leq a<b\leq n-k+1}(\alpha_{b}-\alpha_{a})\neq 0. 
$$
Then $$ \det\begin{pmatrix}
	\boldsymbol{A}_{2}^{(4)}&\boldsymbol{B}_{1}^{(10)}
\end{pmatrix}=\det(\boldsymbol{K}_{2})\det(\boldsymbol{L}_{3})\neq 0,$$
it means $\rank\begin{pmatrix}
	\boldsymbol{A}_{2}^{(4)}&\boldsymbol{B}_{1}^{(10)}
\end{pmatrix}=n$, i.e., 
$$\rank\begin{pmatrix}
	\boldsymbol{G}_{k,*}^{T}&
	\boldsymbol{H}_{n-k,*}^{T}
\end{pmatrix}=\rank\begin{pmatrix}
	\boldsymbol{G}_{k,*}\\
	\boldsymbol{H}_{n-k,*}
\end{pmatrix}=\rank(\boldsymbol{D})=\rank\begin{pmatrix}
	\boldsymbol{A}_{2}^{(4)}&\boldsymbol{B}_{1}^{(10)}
\end{pmatrix}=n,$$
Thus the 
$(*)$-$(\mathcal{L},\mathcal{P})$-$\mathrm{TGRS}$ code is LCD. 

From the above, we complete the proof of Theorem $\ref{LCD3}$.

\subsubsection{The proof of Theorem 4.4}
{\bf Proof}. In the similar proof as that of Theorem \ref{LCD3}, it's enough to prove that  $$\rank\begin{pmatrix}
	\boldsymbol{A}_{2}^{(1)}&\boldsymbol{B}_{1}^{(2)}
\end{pmatrix}=n,$$
where $\boldsymbol{A}_{2}^{(1)}$ and $\boldsymbol{B}_{1}^{(2)}$ are given by $(\ref{LCD1matrixB1(2)})$ and   $(\ref{LCD3matrixA2(1)})$, respectively.

By $n=2k+\ell+r(0\leq r\leq \ell)$, we know $$n-k-(\ell+1)=k+(r-1)\geq k-1,$$
$$n-k-\ell=k+r$$
and
$$n-k=k+\ell+r\geq k+\ell.$$
Then we can carry out the following elementary transformations:

\textbf{Step 1.} Replace $\boldsymbol{a}_{i+1,2}^{(1)}$ with $\boldsymbol{a}_{i+1,2}^{(1)}-\boldsymbol{b}_{i,1}^{(2)}$ for each $i=1,\ldots,k-1$, we can obtain the matrix $\begin{pmatrix}
	\boldsymbol{A}_{2}^{(5)}&\boldsymbol{B}_{1}^{(11)}
\end{pmatrix}$ and denote the vector $\boldsymbol{a}_{i,2}^{(5)}$ and $\boldsymbol{b}_{j,1}^{(11)}$ be the $(i+1)$-th $(0\leq i\leq k-1)$ column and the $j$-th $(1\leq j\leq n-k)$ column of $\boldsymbol{A}_{2}^{(5)}$ and $\boldsymbol{B}_{1}^{(11)}$, respectively; 

\textbf{Step 2.} Replace $\boldsymbol{a}_{1,2}^{(5)}$ with $\boldsymbol{a}_{1,2}^{(5)}-\sum\limits_{t=0}^{\ell}\eta_t\boldsymbol{b}_{k+t,1}^{(11)}$, we can obtain the matrix $\begin{pmatrix}
	\boldsymbol{A}_{2}^{(6)}&\boldsymbol{B}_{1}^{(12)}
\end{pmatrix}$; 

\textbf{Step 3.} put the first column of the matrix $\boldsymbol{A}_{2}^{(6)}$ to the back of the last column of the matrix $\boldsymbol{A}_{2}^{(6)}$, we can obtain the matrix $\begin{pmatrix}
	\boldsymbol{A}_{2}^{(4)}&\boldsymbol{B}_{1}^{(12)}
\end{pmatrix},$
where $\boldsymbol{A}_{2}^{(4)}$ is given by $(\ref{LCD3matrixA2(2)})$ and 
$$
\boldsymbol{B}_{1}^{(12)}=\begin{pmatrix}
	1+\sum\limits_{t=r}^{\ell}\eta_{t}\varPhi_{\ell+1+r-t}&\alpha_{1}&\cdots&\alpha_{1}^{n-k-(\ell+1)}&\alpha_{1}^{n-k-\ell}-\varPhi_{\ell+1}&\cdots&\alpha_{1}^{n-k}-\varPhi_{1}\\ 
	\vdots&\vdots&\ddots&
	\vdots&\vdots&\ddots&\vdots\\
	1+\sum\limits_{t=r}^{\ell}\eta_{t}\varPhi_{\ell+1+r-t}&\alpha_{n-k+1}&\cdots&\alpha_{n-k+1}^{n-k-(\ell+1)}&\alpha_{n-k+1}^{n-k-\ell}-\varPhi_{\ell+1}&\cdots&\alpha_{n-k+1}^{n-k}-\varPhi_{1}\\
	\varXi_{n-k+2}&\alpha_{n-k+2}&\cdots&\alpha_{n-k+2}^{n-k-(\ell+1)}&\alpha_{n-k+2}^{n-k-\ell}-\varPhi_{\ell+1}&\cdots&\alpha_{n-k+2}^{n-k}-\varPhi_{1}\\
	\vdots&\vdots&\ddots&
	\vdots&\vdots&\ddots&\vdots\\
	\varXi_{n}&\alpha_{n}&\cdots&\alpha_{n}^{n-k-(\ell+1)}&\alpha_{n}^{n-k-\ell}-\varPhi_{\ell+1}&\cdots&\alpha_{n}^{n-k}-\varPhi_{1}\\
\end{pmatrix}
$$ 
with $\varXi_{i}=\left(v_{i}^{2}-1\right)\left(1+\sum\limits_{t=0}^{\ell}\eta_{t}\alpha_{i}^{k+t}\right)+1+\sum\limits_{t=r}^{\ell}\eta_{t}\varPhi_{\ell+1+r-t}.$

Note that for any matrix $\boldsymbol{M}$, the rank of $\boldsymbol{M}$ cannot be changed under the elementary transformations  , and so it's enough to prove that  
$$\rank\begin{pmatrix}
	\boldsymbol{A}_{2}^{(4)}&\boldsymbol{B}_{1}^{(12)}
\end{pmatrix}=n,$$
i.e., $\det\begin{pmatrix}
	\boldsymbol{A}_{2}^{(4)}&\boldsymbol{B}_{1}^{(12)}
\end{pmatrix}\neq 0$. For convenience, we denote $\begin{pmatrix}
	\boldsymbol{A}_{2}^{(4)}&\boldsymbol{B}_{1}^{(12)}
\end{pmatrix}\triangleq\begin{pmatrix}
	\boldsymbol{0}&\boldsymbol{L}_{4}\\
	\boldsymbol{K}_{2}&*
\end{pmatrix}$,
where $\boldsymbol{K}_{2}$ is given by $(\ref{LCD3K2})$ and
$$
\small\boldsymbol{L}_{4}=\begin{pmatrix}
	1+\sum\limits_{t=r}^{\ell}\eta_{t}\varPhi_{\ell+1+r-t}&\alpha_{1}& \cdots&\alpha_{1}^{n-k-(\ell +1)}&\alpha_{1}^{n-k-\ell}-\varPhi_{\ell+1}&\cdots&\alpha_{1}^{n-k}-\varPhi_{1}\\ 
	\vdots&\vdots&\ddots&\vdots&\vdots&\ddots&\vdots\\
	1+\sum\limits_{t=r}^{\ell}\eta_{t}\varPhi_{\ell+1+r-t}&\alpha_{n-k+1} &\cdots&\alpha_{n-k+1}^{n-k-(\ell +1)}&\alpha_{n-k+1}^{n-k-\ell}-\varPhi_{\ell+1}&\cdots&\alpha_{n-k+1}^{n-k}-\varPhi_{1}\\
\end{pmatrix}.
$$  
By $(\ref{K2det})$, we have $\det(\boldsymbol{K}_{2})\neq 0.$
And
$$\begin{aligned}
	&\det(\boldsymbol{L}_{4})\\
	=&\left(1+\sum\limits_{t=r}^{\ell}\eta_{t}\varPhi_{\ell+1+r-t}\right)\cdot\det\begin{pmatrix}
		1&\alpha_{1}& \cdots&\alpha_{1}^{n-k-(\ell +1)}&\alpha_{1}^{n-k-\ell}&\cdots&\alpha_{1}^{n-k}\\ 
		\vdots&\vdots&\ddots&\vdots&\vdots&\ddots&\vdots\\
		1&\alpha_{n-k+1} &\cdots&\alpha_{n-k+1}^{n-k-(\ell +1)}&\alpha_{n-k+1}^{n-k-\ell}&\cdots&\alpha_{n-k+1}^{n-k}\\
	\end{pmatrix}\\
	=&\left(1+\sum\limits_{t=r}^{\ell}\eta_{t}\varPhi_{\ell+1+r-t}\right)\cdot\prod\limits_{1\leq a<b\leq n-k+1}(\alpha_{b}-\alpha_{a}).
\end{aligned}
$$
Then by $1+\sum\limits_{t=r}^{\ell}\eta_{t}\varPhi_{\ell+1+r-t}\neq 0$, we can get
$$ \det\begin{pmatrix}
	\boldsymbol{A}_{2}^{(4)}&\boldsymbol{B}_{1}^{(12)}
\end{pmatrix}=\det(\boldsymbol{K}_{2})\det(\boldsymbol{L}_{4})\neq 0,$$
it means $\rank\begin{pmatrix}
	\boldsymbol{A}_{2}^{(4)}&\boldsymbol{B}_{1}^{(12)}
\end{pmatrix}=n$, i.e., 
$$\rank\begin{pmatrix}
	\boldsymbol{G}_{k,*}^{T}&
	\boldsymbol{H}_{n-k,*}^{T}
\end{pmatrix}=\rank\begin{pmatrix}
	\boldsymbol{G}_{k,*}\\
	\boldsymbol{H}_{n-k,*}
\end{pmatrix}=\rank(\boldsymbol{D})=\rank\begin{pmatrix}
	\boldsymbol{A}_{2}^{(4)}&\boldsymbol{B}_{1}^{(12)}
\end{pmatrix}=n,$$
Thus the 
$(*)$-$(\mathcal{L},\mathcal{P})$-$\mathrm{TGRS}$ code is LCD. 

From the above, we complete the proof of Theorem $\ref{LCD4}$.

\subsection{Some examples}
In this sections, we give some examples for Theorems \ref{LCD1}-\ref{LCD4}.

The following Examples \ref{*LPTGRSLCDMDSexample}-\ref{*LPTGRSLCDNMDSexample} are for Theorem \ref{LCD1}.
\begin{example}\label{*LPTGRSLCDMDSexample}	Let $(q,n,k,\lambda)=(61,12,2,1)$. Note that 	$$\begin{aligned}	m(x)=&x^{12}-1\\
	=&(x-1)(x-11)(x-13)(x-14)(x-21)(x-29)\\	&(x-32)(x-40)(x-47)(x-48)(x-50)(x-60).	\end{aligned}$$ Now by taking $\boldsymbol{\alpha}=\left(1, 11, 13, 14, 21, 29, 32, 40, 47, 48, 50, 60\right)\in\mathbb{F}_{61}^{12}$, $ \boldsymbol{\eta}=\left(1,2,3,4\right)\in\mathbb{F}_{61}^{4}\backslash\left\{\boldsymbol{0}\right\}$ and $\boldsymbol{v}=\left(2,1,\ldots,1\right)\in\mathbb{F}_{61}^{12} $, and by directly calculating, we have the following table.	\begin{table}[H]		\centering		\footnotesize 		
		\label{table_example1.3}		\begin{tabular}{|c|c|c|c|c|c|c|c|c|c|c|c|c|}			\hline				$\alpha_{i}$&$1$&$11$&$13$&$14$&$21$&$29$&$32$&$40$&$47$&$48$&$50$&$60$\\			\hline			$\sum\limits_{t=0}^{3}\eta_{t}\alpha_{i}^{2+t}$&$10$&$24$&$32$&$39$&$25$&$7$&$5$&$20$&$25$&$18$&$41$&$59$ \\			\hline			$1+\sum\limits_{t=0}^{3}\eta_{t}\alpha_{i}^{2+t}$&$11$&$25$&$33$&$40$&$26$&$8$&$6$&$21$&$26$&$19$&$42$&$60$ \\			\hline$v_{i}\left(1+\sum\limits_{t=0}^{3}\eta_{t}\alpha_{i}^{2+t}\right)$&$22$&$25$&$33$&$40$&$26$&$8$&$6$&$21$&$26$&$19$&$42$&$60$ \\
			\hline
		\end{tabular}	
	\end{table}
	And then the corresponding code has the generator matrix	$$\boldsymbol{G}_{k,*}=\begin{pmatrix}		22&25&33&40&26&8&6&21&26&19&42&60\\		2&11&13&14&21&29&32&40&47&48&50&60\\	\end{pmatrix}.	$$Furthermore, based on the Magma programe, the corresponding $(*)$-$(\mathcal{L}, \mathcal{P})$-$\mathcal{TGRS}_{k,n}(\boldsymbol{\alpha},\boldsymbol{v},\boldsymbol{B})$ code is MDS with the parameters $\left[12,2,11\right]_{61}.$
\end{example}
\begin{example}\label{*LPTGRSLCDNMDSexample}	Let $(q,n,k,\lambda)=(23,11,2,1)$. Note that 	$$\begin{aligned}		m(x)=x^{11}-1=&(x-1)(x- 2)(x-3)(x- 4)(x-6)(x-8)\\		&(x-9)(x-12)(x-13)(x-16)(x-18).	\end{aligned}$$ Now by taking $\boldsymbol{\alpha}=\left(1, 2, 3, 4, 6, 8, 9, 12, 13, 16, 18\right)\in\mathbb{F}_{23}^{11}$, $ \boldsymbol{\eta}=\left(1,2,3,4\right)\in\mathbb{F}_{23}^{4}\backslash\left\{\boldsymbol{0}\right\}$, $\boldsymbol{v}=\left(2,1,\ldots,1\right)\in\mathbb{F}_{23}^{11} $, and by directly calculating, we have the following table.	
	\begin{table}[H]		\centering		\footnotesize 		
		\label{table_example1.4}		\begin{tabular}{|c|c|c|c|c|c|c|c|c|c|c|c|}			\hline				$\alpha_{i}$&$1$&$2$&$3$&$4$&$6$&$8$&$9$&$12$&$13$&$16$&$18$\\			\hline			$\sum\limits_{t=0}^{3}\eta_{t}\alpha_{i}^{2+t}$&$10$&$12$&$13$&$17$&$17$&$8$&$2$&$8$&$10$&$12$&$6$\\			\hline			$1+\sum\limits_{t=0}^{3}\eta_{t}\alpha_{i}^{2+t}$&$11$&$13$&$14$&$18$&$18$&$9$&$3$&$9$&$11$&$13$&$7$ \\			\hline$v_{i}\left(1+\sum\limits_{t=0}^{3}\eta_{t}\alpha_{i}^{2+t}\right)$&$22$&$13$&$14$&$18$&$18$&$9$&$3$&$9$&$11$&$13$&$7$\\			\hline		\end{tabular}	\end{table}And then the corresponding code has the generator matrix	$$\boldsymbol{G}_{k,*}=\begin{pmatrix}
		22&13&14&18&18&9&3& 9&11&13&7\\		
		2&2&3&4&6&8&9&12&13&16&18\\	\end{pmatrix}.	$$Furthermore, based on the Magma programe, the corresponding $(*)$-$(\mathcal{L}, \mathcal{P})$-$\mathcal{TGRS}_{k,n}(\boldsymbol{\alpha},\boldsymbol{v},\boldsymbol{B})$ code is NMDS with the parameters $\left[11,2,9\right]_{23}.$
\end{example}

The following Examples \ref{*LPTGRSLCD2MDSexample}-\ref{*LPTGRSLCD2NMDSexample} are for Theorem \ref{LCD2}.
\begin{example}\label{*LPTGRSLCD2MDSexample}
	Let $(q,n,k,\lambda,r)=(43,7,2,1,0)$. Note that 	$$m(x)=x^{7}-1=(x-1)(x-4)(x-11)(x-16)(x-21)(x-35)(x-41).$$
	Now by taking $\boldsymbol{\alpha}=\left(1, 4, 11, 16, 21, 35, 41\right)\in\mathbb{F}_{43}^{7}$, $ \boldsymbol{\eta}=\left(1,1,1,1\right)\in\mathbb{F}_{43}^{4}\backslash\left\{\boldsymbol{0}\right\}$, $\boldsymbol{v}=\left(2,1,\ldots,1\right)\in\mathbb{F}_{43}^{7} $, and by directly calculating, we have
	$$P=\prod\limits_{j=1}^{n} \alpha_{j}=(-1)^{n-1}\lambda=1, S_{1}=S_{2}=S_{3}=S_{4}=0$$ 
	and the following table.	\begin{table}[H]		
		\centering		
		\footnotesize 		
		\label{table_example2.1}		\begin{tabular}{|c|c|c|c|c|c|c|c|}		
			\hline				$\alpha_{i}$&$1$&$4$&$11$&$16$&$21$&$35$&$41$\\		
			\hline			$\sum\limits_{t=0}^{3}\eta_{t}\alpha_{i}^{2+t}$&$4$&$27$&$27$&$34$&$23$&$34$&$23$ \\			
			\hline			$1+\sum\limits_{t=0}^{3}\eta_{t}\alpha_{i}^{2+t}$&$5$&$28$&$28$&$35$&$24$&$35$&$24$ \\			\hline$v_{i}\left(1+\sum\limits_{t=0}^{3}\eta_{t}\alpha_{i}^{2+t}\right)$&$10$&$28$&$28$&$35$&$24$&$35$&$24$ \\			
			\hline		
		\end{tabular}	
	\end{table}
	And then the corresponding code has the generator matrix	$$\boldsymbol{G}_{k,*}=\begin{pmatrix}
		10&28&28&35&24&35&24\\		2&4&11&16&21&35&41\\	
	\end{pmatrix}.	
	$$
By directly calculating,  we have
\begin{equation}\label{examplecondition}
\begin{aligned}
		&1+\sum\limits_{t=r}^{\ell}\eta_{t}\varPhi_{\ell+1+r-t}\\
		=&1+\eta_{0}\varPhi_{4}+\eta_{1}\varPhi_{3}+\eta_{2}\varPhi_{2}+\eta_{3}\varPhi_{1}\\
		=&1+P\sum\limits_{t=0}^{3}\eta_{t}S_{t-3}+P\sum\limits_{t=0}^{3}\eta_{t}S_{t-2}+P\sum\limits_{t=0}^{3}\eta_{t}S_{t-1}+P\sum\limits_{t=0}^{3}\eta_{t}S_{t}\\
		=&1+\eta_{3}S_{0}+(\eta_{2}S_{0}+\eta_{3}S_{1})+\left(\eta_{1}S_{0}+\eta_{2}S_{1}+\eta_{3}S_{2}\right)+\left(\eta_{0}S_{0}+\eta_{1}S_{1}+\eta_{2}S_{2}+\eta_{3}S_{3}\right)\\
		=&5\neq 0.
	\end{aligned}
\end{equation}
	Then by Theorem \ref{LCD2}, we know that the corresponding code is LCD. And based on the Magma programe, the corresponding code is LCD MDS with the parameters $\left[7,2,6\right]_{43}.$
\end{example}
\begin{example}\label{*LPTGRSLCD2NMDSexample}
	Let $(q,n,k,\lambda,r)=(41,8,2,1,1)$. Note that 	$$m(x)=x^{8}-1=(x-1)(x-3)(x-9)(x-14)(x-27)(x-32)(x-38)(x-40).$$
	Now by taking $\boldsymbol{\alpha}=\left(1, 3, 9, 14, 27, 32, 38,40\right)\in\mathbb{F}_{41}^{8}$, $ \boldsymbol{\eta}=\left(1,1,1,1\right)\in\mathbb{F}_{41}^{4}\backslash\left\{\boldsymbol{0}\right\}$, $\boldsymbol{v}=\left(2,1,\ldots,1\right)\in\mathbb{F}_{41}^{8} $, and by directly calculating,  we have
	$$P=\prod\limits_{j=1}^{n} \alpha_{j}=-1, S_{1}=S_{2}=0$$
	and the following table.	\begin{table}[H]		
		\centering		
		\footnotesize 		
		\label{table_example2.2}		\begin{tabular}{|c|c|c|c|c|c|c|c|c|}		
			\hline				$\alpha_{i}$&$1$&$3$&$9$&$14$&$27$&$32$&$38$&$40$\\		
			\hline			$\sum\limits_{t=0}^{3}\eta_{t}\alpha_{i}^{2+t}$&$4$&$32$&$0$&$14$&$7$&$0$&$25$&$0$ \\			
			\hline			$1+\sum\limits_{t=0}^{3}\eta_{t}\alpha_{i}^{2+t}$&$5$&$33$&$1$&$15$&$8$&$1$&$26$&$1$ \\			\hline$v_{i}\left(1+\sum\limits_{t=0}^{3}\eta_{t}\alpha_{i}^{2+t}\right)$&$10$&$33$&$1$&$15$&$8$&$1$&$26$&$1$ \\			
			\hline		
		\end{tabular}	
	\end{table}
	And then the corresponding code has the generator matrix	$$\boldsymbol{G}_{k,*}=\begin{pmatrix}
		10&33&1&15&8&1&26&1\\
		2&3&9&14&27&32&38&40	
	\end{pmatrix}.	
	$$
In the similar calculation method as that of $(\ref{examplecondition})$, we have
$$1+\sum\limits_{t=r}^{\ell}\eta_{t}\varPhi_{\ell+1+r-t}=-2\neq0.$$	
	Then by Theorem \ref{LCD2}, we know that the corresponding code is LCD. And based on the Magma programe, the corresponding code is LCD NMDS with the parameters $\left[8,2,6\right]_{41}.$
\end{example}

The following Examples \ref{*LPTGRSLCD3MDSexample}-\ref{*LPTGRSLCD3NMDSexample} are for Theorem \ref{LCD3}.
\begin{example}\label{*LPTGRSLCD3MDSexample}	Let $(q,n,k,\lambda)=(61,12,2,1)$. Note that 	$$\begin{aligned}	m(x)=x^{12}-1=&(x-1)(x-11)(x-13)(x-14)(x-21)(x-29)\\	&(x-32)(x-40)(x-47)(x-48)(x-50)(x-60).	\end{aligned}$$
Now by taking $\boldsymbol{\alpha}=\left(1, 11, 13, 14, 21, 29, 32, 40, 47, 48, 50, 60\right)\in\mathbb{F}_{61}^{12}$, $ \boldsymbol{\eta}=\left(1,2,3,4\right)\in\mathbb{F}_{61}^{4}\backslash\left\{\boldsymbol{0}\right\}$, $\boldsymbol{v}=\left(-1, 1, -1, -1, -1, 1, 1, 1, 1, 1, -1, 2\right)\in\mathbb{F}_{61}^{12},$ and by directly calculating, we have the following table.	\begin{table}[H]		\centering		\footnotesize 		
		\label{table_example3.1}		\begin{tabular}{|c|c|c|c|c|c|c|c|c|c|c|c|c|}			\hline				$\alpha_{i}$&$1$&$11$&$13$&$14$&$21$&$29$&$32$&$40$&$47$&$48$&$50$&$60$\\			\hline			$\sum\limits_{t=0}^{3}\eta_{t}\alpha_{i}^{2+t}$&$10$&$24$&$32$&$39$&$25$&$7$&$5$&$20$&$25$&$18$&$41$&$59$ \\			\hline			$1+\sum\limits_{t=0}^{3}\eta_{t}\alpha_{i}^{2+t}$&$11$&$25$&$33$&$40$&$26$&$8$&$6$&$21$&$26$&$19$&$42$&$60$ \\			\hline$v_{i}\left(1+\sum\limits_{t=0}^{3}\eta_{t}\alpha_{i}^{2+t}\right)$&$50$&$25$&$28$&$21$&$35$&$8$&$6$&$21$&$26$&$19$&$19$&$59$ \\
			\hline
		\end{tabular}	
	\end{table}
	And then the corresponding code has the generator matrix	$$\boldsymbol{G}_{k,*}=\begin{pmatrix}
	50&25&28&21&35&8&6&21&26&19&19&59\\
	60&11&48&47&40&29&32&40&47&48&11&59
	\end{pmatrix}.$$
Then by Theorem \ref{LCD3}, we know that the corresponding code is LCD. And based on the Magma programe, the corresponding code is LCD MDS with the parameters $\left[12,2,11\right]_{61}.$
\end{example}

\begin{example}\label{*LPTGRSLCD3NMDSexample}	Let $(q,n,k,\lambda)=(23,11,2,1)$. Note that 	$$\begin{aligned}		m(x)=x^{11}-1=&(x-1)(x- 2)(x-3)(x- 4)(x-6)(x-8)\\		&(x-9)(x-12)(x-13)(x-16)(x-18).	\end{aligned}$$Now by taking $\boldsymbol{\alpha}=\left(1, 2, 3, 4, 6, 8, 9, 12, 13, 16, 18\right)\in\mathbb{F}_{23}^{11}$, $ \boldsymbol{\eta}=\left(1,2,3,4\right)\in\mathbb{F}_{23}^{4}\backslash\left\{\boldsymbol{0}\right\}$, $\boldsymbol{v}=\left(-1, 1, -1, -1, -1, -1, 1, -1, -1, -1, 2\right)\in\mathbb{F}_{23}^{11} $, and by directly calculating, we have the following table.	
	\begin{table}[H]		\centering		\footnotesize 		
		\label{table_example3.2}		\begin{tabular}{|c|c|c|c|c|c|c|c|c|c|c|c|}			\hline				$\alpha_{i}$&$1$&$2$&$3$&$4$&$6$&$8$&$9$&$12$&$13$&$16$&$18$\\			\hline			$\sum\limits_{t=0}^{3}\eta_{t}\alpha_{i}^{2+t}$&$10$&$12$&$13$&$17$&$17$&$8$&$2$&$8$&$10$&$12$&$6$\\			\hline			$1+\sum\limits_{t=0}^{3}\eta_{t}\alpha_{i}^{2+t}$&$11$&$13$&$14$&$18$&$18$&$9$&$3$&$9$&$11$&$13$&$7$ \\			\hline$v_{i}\left(1+\sum\limits_{t=0}^{3}\eta_{t}\alpha_{i}^{2+t}\right)$&$12$&$13$&$9$&$5$&$5$&$14$&$3$&$14$&$12$&$10$&$14$\\			\hline		\end{tabular}	\end{table}And then the corresponding code has the generator matrix	$$\boldsymbol{G}_{k,*}=\begin{pmatrix}
		12&13&9&5&5&14&3&14&12&10&14\\
		22&2&20&19&17&15&9&11&10&7&13	\end{pmatrix}.	$$
	Then by Theorem \ref{LCD3}, we know that the corresponding code is LCD. And based on the Magma programe, the corresponding code is LCD NMDS with the parameters $\left[11,2,9\right]_{23}.$
\end{example}

The following Examples \ref{*LPTGRSLCD4MDSexample}-\ref{*LPTGRSLCD4NMDSexample} are for Theorem \ref{LCD4}.
\begin{example}\label{*LPTGRSLCD4MDSexample}
	Let $(q,n,k,\lambda,r)=(73,9,2,1,1)$. Note that 	$$m(x)=x^{9}-1=(x-1)(x-2)(x-4)(x-4)(x-8)(x-16)(x-32)(x-37)(x-55)(x-64)\in\mathbb{F}_{73}[x].$$
	Now by taking $\boldsymbol{\alpha}=\left(1, 2, 4, 8, 16, 32, 37, 55, 64\right)\in\mathbb{F}_{73}^{9}$, $ \boldsymbol{\eta}=\left(1,2,3,4\right)\in\mathbb{F}_{73}^{4}\backslash\left\{\boldsymbol{0}\right\}$, $\boldsymbol{v}=\left(1, 1, -1, -1, -1, 1, 1, -1, 2\right)\in\mathbb{F}_{73}^{9} $ and by directly calculating,  we have
	$$P=\prod\limits_{i=1}^{n}\alpha_{i}=1, S_{1}=S_{2}=0.$$ and the following table.	\begin{table}[H]		
	 	\centering		
	 	\footnotesize 		
	 	\label{table_example4.1}		\begin{tabular}{|c|c|c|c|c|c|c|c|c|c|}		
	 		\hline				$\alpha_{i}$&$1$&$2$&$4$&$8$&$16$&$32$&$37$&$55$&$64$\\		
	 		\hline			$\sum\limits_{t=0}^{3}\eta_{t}\alpha_{i}^{2+t}$&$10$&$50$&$44$&$54$&$15$&$70$&$51$&$56$&$15$ \\			
	 		\hline			$1+\sum\limits_{t=0}^{3}\eta_{t}\alpha_{i}^{2+t}$&$11$&$51$&$45$&$55$&$16$&$71$&$52$&$57$&$16$ \\			\hline$v_{i}\left(1+\sum\limits_{t=0}^{3}\eta_{t}\alpha_{i}^{2+t}\right)$&$11$&$51$&$28$&$18$&$57$&$71$&$52$&$16$&$32$\\			
	 		\hline		
	 	\end{tabular}	
	 \end{table}
	 And then the corresponding code has the generator matrix	$$\boldsymbol{G}_{k,*}=\begin{pmatrix}
	 	11&51&28&18&57&71&52&16&32\\
	 	1&2&69&65&57&32&37&18&55
	 \end{pmatrix}.
	 $$
In the similar calculation method as that of $(\ref{examplecondition})$, we have
$$1+\sum\limits_{t=r}^{\ell}\eta_{t}\varPhi_{\ell+1+r-t}
=21\neq0.$$
	Then by Theorem \ref{LCD4}, we know that the corresponding code is LCD. And based on the Magma programe, the corresponding code is LCD MDS with the parameters $\left[9,2,8\right]_{73}.$
\end{example}
\begin{example}\label{*LPTGRSLCD4NMDSexample}
	Let $(q,n,k,\lambda,r)=(29,7,2,1,0)$. Note that 	$$m(x)=x^{7}-1=(x-1)(x-7)(x-16)(x-20)(x-23)(x-24)(x-25)\in\mathbb{F}_{29}[x].$$
	Now by taking $\boldsymbol{\alpha}=\left(1, 7, 16, 20, 23, 24, 25\right)\in\mathbb{F}_{29}^{7}$, $ \boldsymbol{\eta}=\left(1,1,1,1\right)\in\mathbb{F}_{29}^{4}\backslash\left\{\boldsymbol{0}\right\}$, $\boldsymbol{v}=\left(1,-1, -1, 1, 1,-1, 2\right)\in\mathbb{F}_{29}^{7} $ and by directly calculating,  we have $$P=\prod\limits_{i=1}^{n}\alpha_{i}=1, S_{1}=\sum\limits_{i=1}\alpha_{i}=0$$ and the following table.	\begin{table}[H]		
	\centering		
	\footnotesize 		
	\label{table_example4.2}		\begin{tabular}{|c|c|c|c|c|c|c|c|}		
		\hline				$\alpha_{i}$&$1$&$7$&$16$&$20$&$23$&$24$&$25$\\		
		\hline			$\sum\limits_{t=0}^{3}\eta_{t}\alpha_{i}^{2+t}$&$4$&$25$&$21$&$21$&$10$&$10$&$25$\\			
		\hline			$1+\sum\limits_{t=0}^{3}\eta_{t}\alpha_{i}^{2+t}$&$5$&$26$&$22$&$22$&$11$&$11$&$26$ \\			\hline$v_{i}\left(1+\sum\limits_{t=0}^{3}\eta_{t}\alpha_{i}^{2+t}\right)$&$5$&$3$&$7$&$22$&$11$&$18$&$23$\\			
		\hline		
	\end{tabular}	
\end{table}
And then the corresponding code has the generator matrix	$$\boldsymbol{G}_{k,*}=\begin{pmatrix}
5&3&7&22&11&18&23\\
1&22&13&20&23&5&21
\end{pmatrix}.	
$$
In the similar calculation method as that of $(\ref{examplecondition})$, we have
$$1+\sum\limits_{t=r}^{\ell}\eta_{t}\varPhi_{\ell+1+r-t}\\
=5\neq 0.$$
	Then by Theorem \ref{LCD4}, we know that the corresponding code is LCD. And based on the Magma programe, the corresponding code is LCD NMDS with the parameters $\left[7,2,5\right]_{29}.$
\end{example} 

\section{Conclusion}
In this paper, we define the  $(*)$-$(\mathcal{L}, \mathcal{P})$-$\mathrm{TGRS}_{k,n}(\boldsymbol{\alpha},\boldsymbol{v},\boldsymbol{\eta})$ code and obtain the following two main results.

\begin{itemize}
\item A parity-check matrix of the $(*)$-$(\mathcal{L}, \mathcal{P})$-$\mathrm{TGRS}_{k,n}(\boldsymbol{\alpha},\boldsymbol{v},\boldsymbol{\eta})$ code (Theorem $\ref{*LPTGRSparitymatrix}$).

\item Four classes of LCD $(*)$-$(\mathcal{L}, \mathcal{P})$-$\mathrm{TGRS}_{k,n}(\boldsymbol{\alpha},\boldsymbol{v},\boldsymbol{\eta})$ codes (Theorems $\ref{LCD1}$-$\ref{LCD4}$).
\end{itemize}
Furthermore, our results are the generalization for the corresponding results in \cite{A19} and \cite{A21}. The detail can be seen the following  table:
\begin{table}[H]
\centering 
\label{table_example7.1}
\renewcommand{\arraystretch}{1.5}
\begin{tabular}{|c|c|c|c|c|}
\hline
Our results&$\ell$&$\boldsymbol{\eta}= \left(\eta_{0}, \ldots, \eta_{\ell}\right) \in\mathbb{F}_{q}^{\ell+1}\backslash\left\{\boldsymbol{0}\right\}$&$r$& Known results \\		\hline 
\multirow{3}{*}{Theorem \ref{*LPTGRSparitymatrix}}  &$0$&$\eta_{0}\in\mathbb{F}_{q}^{*}$& &\cite{A19} Theorem 1\\	
\cline{2-5}
 &$1$&$\left(\eta_{0},\eta_{1}\right)\in\left(\mathbb{F}_{q}^{*}\right)^{2}$& &\cite{A21} Lemma 2.8 \\	
 \cline{2-5}
 &$2$&$\left(\eta_{0},\eta_{1},\eta_{2}\right)\in\left(\mathbb{F}_{q}^{*}\right)^{3},\eta_{1}=0$& &\cite{A21} Lemma 2.9\\	
\hline 
\multirow{3}{*}{Theorem \ref{LCD1}}&$0$&$\eta_{0}\in\mathbb{F}_{q}^{*}$& &\cite{A19} Theorem 2 \\	
\cline{2-5} 
 &$1$&$\left(\eta_{0},\eta_{1}\right)\in\left(\mathbb{F}_{q}^{*}\right)^{2}$& &\cite{A21} Theorem 5.1 \\	
\cline{2-5}
&$2$&$\left(\eta_{0},\eta_{1},\eta_{2}\right)\in\left(\mathbb{F}_{q}^{*}\right)^{3},\eta_{1}=0$& &\cite{A21} Theorem 5.4\\	
\hline 
\multirow{6}{*}{Theorem \ref{LCD2}}&$0$&$\eta_{0}\in\mathbb{F}_{q}^{*}$&$0$&\cite{A19} Theorem 2 \\
\cline{2-5}	  
&\multirow{2}{*}{$1$}&\multirow{2}{*}{$ \left(\eta_{0},\eta_{1}\right)\left(\mathbb{F}_{q}^{*}\right)^{2}$}&$1$&\cite{A21} Theorem 5.7 \\	
\cline{4-5}
& & &$0$&\cite{A21} Theorem 5.18 \\	
\cline{2-5}
&\multirow{3}{*}{$2$}&\multirow{3}{*}{$\left(\eta_{0},\eta_{1},\eta_{2}\right)\left(\mathbb{F}_{q}^{*}\right)^{3},\eta_{1}=0$}&$0$&\cite{A21} Theorem 5.9\\	
\cline{4-5}
&&&$1$&\cite{A21} Theorem 5.12\\
\cline{4-5}
&&&$0$&\cite{A21} Theorem 5.20\\
\hline   
\multirow{2}{*}{Theorem \ref{LCD3}}&$1$&$\left(\eta_{0},\eta_{1}\right)\left(\mathbb{F}_{q}^{*}\right)^{2}$& &\cite{A21} Theorem 5.14 \\	
\cline{2-5}	 
&$2$& $\left(\eta_{0},\eta_{1},\eta_{2}\right)\left(\mathbb{F}_{q}^{*}\right)^{3},\eta_{1}=0$&&\cite{A21} Theorem 5.16\\
\hline     
		\end{tabular}
	\end{table}

\begin{thebibliography}{100}

\bibitem{A1} Abdukhalikov K, Ding C, Verma G K. Some constructions of non-generalized Reed-Solomon MDS Codes[J]. arXiv preprint arXiv:2506.04080, 2025.
\bibitem{A2} Liu W, Luo J, Wang P, et al. Column Twisted Reed-Solomon Codes as MDS Codes[J]. arXiv preprint arXiv:2507.08755, 2025.
\bibitem{A3} Chen H. Many non-Reed-Solomon type MDS codes from arbitrary genus algebraic curves[J]. IEEE Transactions on Information Theory, 2023, 70(7): 4856-4864.
\bibitem{A4} Y. Wu Y, Heng Z, Li C, et al. More MDS codes of non-Reed-Solomon type[J]. arXiv preprint arXiv:2401.03391, 2024.
\bibitem{A5} Liu S, Liu H, Chen B. Construction of non-generalized Reed-Solomon MDS codes based on systematic generator matrix[J]. arXiv preprint arXiv:2507.20559, 2025.
\bibitem{A6} Liang Z, Liao Q. The equivalent condition for GRL codes to be MDS, AMDS or self-dual[J]. arXiv preprint arXiv:2506.03874, 2025.
\bibitem{A7} Beelen P, Puchinger S, né Nielsen J R. Twisted reed-solomon codes[C]//2017 IEEE International Symposium on Information Theory (ISIT). IEEE, 2017: 336-340.
\bibitem{A8} Beelen P, Puchinger S, Rosenkilde J. Twisted Reed–Solomon codes[J]. IEEE transactions on information theory, 2022, 68(5): 3047-3061.
\bibitem{A9} Wu Y. Twisted Reed–Solomon codes with one-dimensional hull[J]. IEEE Communications Letters, 2020, 25(2): 383-386.
\bibitem{A10} Huang D, Yue Q, Niu Y, et al. MDS or NMDS self-dual codes from twisted generalized Reed–Solomon codes[J]. Designs, Codes and Cryptography, 2021, 89(9): 2195-2209.
\bibitem{A11} Sui J, Yue Q, Li X, et al. MDS, near-MDS or 2-MDS self-dual codes via twisted generalized Reed-Solomon codes[J]. IEEE Transactions on Information Theory, 2022, 68(12): 7832-7841.
\bibitem{A12} Gu H, Zhang J. On twisted generalized Reed-Solomon codes with $\ell$ twists[J]. IEEE Transactions on Information Theory, 2023, 70(1): 145-153.
\bibitem{A13} Sun H, Yue Q, Jia X, et al. Decoding algorithms of twisted GRS codes and twisted Goppa codes[J]. IEEE Transactions on Information Theory, 2024.
\bibitem{A14} Zhu C, Liao Q. A class of double-twisted generalized Reed-Solomon codes[J]. Finite Fields and Their Applications, 2024, 95: 102395. 
\bibitem{A15} Zhao C, Ma W, Yan T, et al. Research on the Construction of Maximum Distance Separable Codes via Arbitrary Twisted Generalized Reed-Solomon Codes[J]. IEEE Transactions on Information Theory, 2025.
\bibitem{A16} Hu Z, Wang L, Li N, et al. On $(\mathcal {L},\mathcal {P}) $-Twisted Generalized Reed-Solomon Codes[J]. arXiv preprint arXiv:2502.04746, 2025.
\bibitem{A17} Carlet C, Guilley S. Complementary Dual Codes for Counter-Measures to Side-Channel Attacks[C]//ICMCTA. 2014: 97-105.
\bibitem{A18} Jin L. Construction of MDS codes with complementary duals[J]. IEEE Transactions on Information Theory, 2016, 63(5): 2843-2847.
\bibitem{A19} Chen B, Liu H. New constructions of MDS codes with complementary duals[J]. IEEE Transactions on Information Theory, 2017, 64(8): 5776-5782.
\bibitem{A20} Carlet C, Mesnager S, Tang C, et al. Euclidean and Hermitian LCD MDS codes[J]. Designs, Codes and Cryptography, 2018, 86(11): 2605-26. 
\bibitem{A21} Beelen P, Jin L. Explicit MDS codes with complementary duals[J]. IEEE Transactions on Information Theory, 2018, 64(11): 7188-7193.
\bibitem{A22} Shi X, Yue Q, Yang S. New LCD MDS codes constructed from generalized Reed–Solomon codes[J]. Journal of Algebra and Its Applications, 2019, 18(08): 1950150.  
\bibitem{A23} Wu Y, Hyun J Y, Lee Y. New LCD MDS codes of non-Reed-Solomon type[J]. IEEE Transactions on Information Theory, 2021, 67(8): 5069-5078.
\bibitem{A24} Liu H, Liu S. Construction of MDS twisted Reed–Solomon codes and LCD MDS codes[J]. Designs, Codes and Cryptography, 2021, 89(9): 2051-2065.
\bibitem{A25} Huang D, Yue Q, Niu Y. MDS or NMDS LCD codes from twisted Reed-Solomon codes[J]. Cryptography and communications, 2023, 15(2): 221-237. 
\bibitem{A26} Zhu C, Liao Q. The [1, 0]-twisted generalized Reed-Solomon code[J]. Cryptography and Communications, 2024, 16(4): 857-878.
\bibitem{A27} Yang S, Wang J, Wu Y. Two classes of twisted generalized Reed-Solomon codes with two twists[J]. Finite Fields and Their Applications, 2025, 104: 102595.
\bibitem{A28} Wan D, Zhang J. Complete symmetric polynomials over finite fields have many rational zeros (in Chinese). Sci
Sin Math, 2021, 51: 1677–1684, doi: 10.1360/SSM-2020-0328.
\end{thebibliography}
\end{document}